\documentclass[preprint,showpacs,preprintnumbers,amsmath,amssymb,superscriptaddress,nofootinbib]{revtex4}
\usepackage{graphicx,color}
\usepackage{amsmath,amssymb}
\usepackage{url}
\usepackage{epstopdf}

\newcommand{\be}{\begin{equation}}
\newcommand{\ee}{\end{equation}}
\newcommand{\bea}{\begin{eqnarray}}
\newcommand{\eea}{\end{eqnarray}}

\newcommand{\crn}{\nonumber \\}

\newcommand{\al}{\alpha}

\newcommand{\bc}{\begin{center}}
\newcommand{\ec}{\end{center}}

\newcommand{\de}{\delta}
\newcommand{\De}{\Delta}

\newcommand {\ba}{\begin{array}}
\newcommand {\ea}{\end{array}}
\newcommand{\ben}{\begin{enumerate}}
\newcommand{\een}{\end{enumerate}}

\usepackage{bm}
\usepackage{dcolumn}

\begin{document}

\title{ Large $(g-2)_{\mu}$ and signals of  decays  $e_b\rightarrow e_a\gamma$  in a 3-3-1 model with inverse seesaw neutrinos }

\author{L.~T.~Hue}\email{lethohue@duytan.edu.vn}
\affiliation{Institute for Research and Development, Duy Tan University, Da Nang City 50000, Vietnam}

\author{H.~T.~Hung}\email{hthung80@gmail.com}
\affiliation{Department of Physics, Hanoi Pedagogical University 2, Phuc Yen, Vinh Phuc 15000, Vietnam}
\author{N.~T.~Tham}\email{nguyenthitham@hpu2.edu.vn}
\affiliation{Department of Physics, Hanoi Pedagogical University 2, Phuc Yen, Vinh Phuc 15000, Vietnam}

\author{H. N. Long}\email{hnlong@iop.vast.ac.vn}
\affiliation{Institute of Physics,   Vietnam Academy of Science and Technology, 10 Dao Tan, Ba Dinh, 10000 Hanoi, Vietnam}

\author{T.Phong Nguyen \footnote{ Corresponding author}}\email{thanhphong@ctu.edu.vn }
\affiliation{Department of Physics, Can Tho University, 3/2 Street, Ninh Kieu, Can Tho City 94000, Vietnam}

\date{\today }
\begin{abstract}
We  show that  under current experimental bounds of the decays $e_a\rightarrow e_b\gamma$, the recent experimental data of the muon anomalous magnetic dipole moment  $(g-2)_{\mu}$ can be explained in the framework of the 3-3-1 model with right-handed neutrinos. In addition, all of these  branching ratios  can reach closely  the recent experimental upper bounds.

\end{abstract}
\maketitle
\allowdisplaybreaks
 \section{\label{intro} Introduction}

At present, the experimental data on the anomalous dipole
magnetic moments  of electron and muon $a_{e,\mu }=(g_{e,\mu }-2)/2$ show
significant deviations from their values predicted by the Standard Model (SM) \cite{Hagiwara:2011af, Davier:2017zfy, Parker:2018vye, Zyla:2020zbs}.
 From the combination of various different contributions~\cite{Davier:2017zfy, Keshavarzi:2018mgv, Colangelo:2018mtw, Hoferichter:2019mqg, Davier:2019can, Keshavarzi:2019abf, Kurz:2014wya, Melnikov:2003xd, Masjuan:2017tvw, Colangelo:2017fiz, Hoferichter:2018kwz, Gerardin:2019vio, Bijnens:2019ghy, Colangelo:2019uex, Colangelo:2014qya, Blum:2019ugy, Aoyama:2012wk, Aoyama:2019ryr, Czarnecki:2002nt, Gnendiger:2013pva}, the
 recent improved value of $a_{\mu}$ predicted by the SM 
  is  accepted widely as follows~\cite{Aoyama:2020ynm}: $a^{\mathrm{SM}}_{\mu}= 116591810(43)\times 10^{-11}$.   The latest experimental measurement  has been reported from  Fermi National Accelerator Laboratory~\cite{Abi:2021gix},  $a^{\mathrm{exp}}_{\mu}=116592061(41)\times 10^{-11}$, leading to the   improved standard deviation of 4.2 $\sigma$ from the SM prediction, namely
 \be\label{eq_damu}
 \Delta a_{\mu}\equiv  a^{\mathrm{exp}}_{\mu} -a^{\mathrm{SM}}_{\mu} =251 \times 10^{-11}\pm 59 \times 10^{-11}.
 \ee

On the other hand, the recent constraints on the charged lepton flavor violating  (cLFV) decays, $e_b\rightarrow e_a\gamma$ are~\cite{TheMEG:2016wtm, Aubert:2009ag}:
 \begin{align}
 \label{eq_ebagaex}
 \mathrm{Br}(\tau\rightarrow \mu\gamma)&<4.4\times 10^{-8},\crn
 \mathrm{Br}(\tau\rightarrow e\gamma)&<3.3\times 10^{-8},\crn
 \mathrm{Br}(\mu\rightarrow e\gamma)&< 4.2\times 10^{-13}.
 \end{align}
 Many recent versions of the  3-3-1 models were indicated that they are difficult to  explain simultaneously all  of these experimental  constraints~\cite{ Kelso:2014qka, Binh:2015jfz, Lindner:2016bgg, deJesus:2020upp, deJesus:2020ngn, Ky:2000ku} with the very large TeV values of the  $SU(3)_L$ symmetry scale.  Namely, the discussion on Ref.~\cite{Kelso:2014qka}  needs the cLFV constraints from experimental data to rule out large $\Delta a_{\mu}$. The remaining models  rule out large $\Delta a_{\mu}$  for large  $SU(3)_L$ symmetry scale with order of $\mathcal{O}(1)$ TeV, if no new $SU(3)_L$ Higgs triplet or vectorlike charged lepton are added.   This result can be explained qualitatively from a consequence that a one-loop contribution from a  heavy gauge boson $V$ is different from that of the  $W^{\pm}$ boson  by a small factor $m^2_W/m_V^2 \ge 10^{-3}$.  Similarly, one-loop contributions from heavy Higgs boson $S$ have a suppressed factor $m^2_{h}/m^2_S$, where $m_{h}$ is the mass of the  standard model (SM-like) Higgs boson. In addition, these  Higgs contributions are constrained strictly by the small upper bound of Br$(\mu\rightarrow e\gamma)$, leading to a strict constraint on the doubly Higgs mass for the 3-3-1 models adding a $SU(3)_L$ Higgs sextet to explain the experimental neutrino oscillation data. Adding  new particles as Higgs triplets or vectorlike charged leptons  into the original 3-3-1 models to generate new couplings contributing to $\Delta a_{\mu}$ is a popular way to explain successful the experimental data of $a_{\mu}$  \cite{deJesus:2020upp, deJesus:2020ngn}, but there seems irrelevant  with neutrino oscillation data. Some recent extensions of 3-3-1 models with discrete symmetries~\cite{CarcamoHernandez:2019lhv, CarcamoHernandez:2020pxw}  need a  large number of new leptons and Higgs bosons for  the explanation of large $\Delta a_{\mu,e}$ consistent with experiments. On the other hand,  a recent note indicated that  a version of the 3-3-1 model with right-handed neutrino (331RN) with heavy neutral fermions assigned as $SU(3)_L$ gauge singlets (called the 331ISS model for short) can predict large one-loop contributions from singly charged Higgs bosons  and inverse seesaw (ISS) neutrinos enough to explain the recent $(g-2)_{\mu}$ data~\cite{Dinh:2020pqn}. More interesting, the model contains two  singly charged Higgs bosons, which may result in a special possibility that  two one-loop contributions to $\De a_{\mu}$  are large and constructive, while those   relate with cLFV decay amplitudes are strongly destructive. In this work, we will pay attention to this possibility,  namely we will  try to answer a question  whether there exist any allowed regions  of the parameter space that the destructive properties of the Higgs contributions are enough to satisfy the cLFV experimental constraints given in Eq.~\eqref{eq_ebagaex}, and explain successfully the recent data given in Eq.~\eqref{eq_damu}.  We will use the 3-3-1 model with the general Higgs potential given in Ref.~\cite{Chang:2006aa, Sanchez-Vega:2016dwe}.
The 3-3-1 models explaining active neutrino data based on the  ISS mechanism has been discussed widely previously~\cite{Catano:2012kw, Dias:2012xp, Pires:2018kaj, Boucenna:2015zwa}, but the interesting regions of the parameter space allowing large $\Delta a_{\mu}$ data and consistent with recent cLFV experimental constraints  were not shown.  In addition, the Br$(\tau\rightarrow\mu\gamma,e\gamma)$ were predicted to be smaller than Br$(\mu\rightarrow e\gamma)$, which is very suppressed with the recent and upcoming experimental sensitivities of the order $\mathcal{O}(10^{-9})$~\cite{Baldini:2013ke, Aushev:2010bq}. Many other models  beyond the SM with the ISS mechanism can explain consistently the experimental data of  $\De a_{\mu}$ and cLFV constraints  ~\cite{Cao:2019evo, Cao:2021lmj, Nomura:2021adf, Mondal:2021vou}.  Here we analyze predictions of the 3-3-1  model with right-handed neutrinos  for the above observables.

Our work is arranged as follows. We will review the 331ISS model in Sec.~\ref{sec_model},  summarize  the gauge, Higgs  bosons and the lepton sectors.  In Sec.~\ref{sec_amuFormula}, we introduce the analytic formulas to calculate the muon magnetic dipole moment and the cLFV branching ratios. In Sec.~\ref{sec_HiggSinglet}, we discuss on the effect of a new singly charged Higgs boson that can give one-loop contributions to $\De a_{\mu}$ and cLFV amplitudes enough to explain successful all the experimental data under consideration. In Sec.~\ref{sec_numerical}, illustrations for numerical results are given to indicate the existence of the allowed regions satisfying the experimental data mentioned in this work. The conclusion is presented in the last  Sec.~\ref{sec:conclusion}, where important results will be summarized.

\section{\label{sec_model} Review the 3-3-1ISS model}
\subsection{Gauge bosons  and  fermions }
The particle content of the 331ISS model was introduced in Refs.~\cite{Boucenna:2015zwa, Nguyen:2018rlb} where  active neutrino masses and oscillations are  originated from the ISS mechanism. The quark sector and $SU(3)_C$ representations are irrelevant in this work, and hence they are omitted here.  We refer Ref.~\cite{Boucenna:2015zwa} for a quark discussion.  The electric charge operator corresponding to  the gauge group  $SU(3)_L\times U(1)_X$ is $Q=T_3-\frac{1}{\sqrt{3}}T_8+X$, where $T_{3,8}$ are the  diagonal $SU(3)_L$ generators. Each  lepton family consists of  a  $SU(3)_L$ triplet $\psi_{aL}= (\nu_a,~e_a, N_a)_L^T\sim (3,-\frac{1}{3})$ and a right-handed charged lepton  $e_{aR}\sim (1,-1)$ with $a=1,2,3$.  Each left-handed neutrino $N_{aL}=(N_{aR})^c$  is equivalent with  a new right-handed neutrinos defined in previous 331RN models~\cite{Foot:1994ym}.  The only difference  between the  two models  331RN and 331ISS is that,  the 331ISS model contains  three more right-handed neutrinos transforming as gauge singlets, $X_{aR}\sim (1,0)$, $a=1,2,3$. They couple with the $SU(3)_L$ Higgs triplets to generate the neutrino mass term relating with the ISS mechanism. The three Higgs triplets  $\rho=(\rho^+_1,~\rho^0,~\rho^+_2)^T\sim (3,\frac{2}{3})$,
$\eta=(\eta_1^0,~\eta^-,\eta^0_2)^T\sim (3,-\frac{1}{3})$, and $\chi=(\chi_1^0,~\chi^-,\chi^0_2)^T\sim (3,-\frac{1}{3})$ have the following   necessary vacuum expectation values for generating all tree-level quark masses and leptons: $\langle\rho \rangle=(0,\,\frac{v_1}{\sqrt{2}},\,0)^T$, $\langle \eta \rangle=(\frac{v_2}{\sqrt{2}},\,0,\,0)^T$ and $\langle \chi \rangle=(0,\,0,\,\frac{w}{\sqrt{2}})^T$.

The gauge bosons  get masses through the covariant kinetic term of the Higgs triplets, $\mathcal{L}^{H}=\sum_{H=\chi,\eta,\rho} \left(D_{\mu}H\right)^{\dagger}\left(D^{\mu}H\right)$, where the covariant derivative for the electroweak symmetry is
$D_\mu  = \partial _\mu  - i g  {W}_\mu ^a{T^a} - i{g_X}{T^9}X{X_\mu }$,  $a=1,2,..,8$. Note that   $T^9 \equiv \frac{I_3}{\sqrt{6}}$ and $\frac{1}{\sqrt{6}}$ for (anti)triplets and singlets \cite{Buras:2012dp}. Matching with the SM gives  $e=g\, s_W$ and $ \frac{g_X}{g}= \frac{3\sqrt{2}s_W}{\sqrt{3-4s^2_W}}$,
where $e$ and $s_W$ are  respective the electric charge and sine of the Weinberg angle, $s^2_W\simeq 0.231$. The relation $\frac{g_X}{g}$  is the same for both choices of triplet or antitriplets representations of the left-handed leptons~\cite{Buras:2014yna, Hue:2018dqf}. The derivation of this  relation is summarized as follows.  The 3-3-1 models have two spontaneous breaking steps: $SU(3)_L\times U(1)_X\overset{w}{\longrightarrow}   SU(2)_L\times U(1)_Y \overset{v_1,v_2}{\longrightarrow}  U(1)_Q$. The first breaking step with $w\neq0$ generates masses for heavy particles predicted by the $SU(3)_L$ symmetry.  The neutral gauge bosons will change into the basis containing the SM ones $W^3_{\mu}$ and $B_{\mu}$:   $(W^3_{\mu},\; W^{8}_{\mu}, X_{\mu})\overset{w\neq0, v_1=v_2=0}{\longrightarrow} (W^3_{\mu}, Z'_{\mu}, B_{\mu})$. Diagonalizing the squared mass matrix of these neutral gauge bosons will get a massive eigenstate $Z'$ with $m^2_{Z'}\sim w^2$ and two SM massless states $W^3_\mu$ and $B_{\mu}$. The relations between the two bases before and after the first breaking step are $W^8_{\mu}=\frac{\beta t}{\sqrt{6+\beta^2 t^2}} B_{\mu}  - \frac{\sqrt{6}}{\sqrt{6+\beta^2 t^2}} Z'_{\mu}$, and $X_{\mu}=\frac{\sqrt{6}}{\sqrt{6+\beta^2 t^2}}  B_{\mu}   +\frac{\beta t}{\sqrt{6+\beta^2 t^2}}Z'_{\mu}$, with $t\equiv g_X/g$. Inserting these  relations to the covariant derivation of the $3-3-1$ gauge group and keeping the part used to identify with the SM one, we have
\begin{align*}
D^{3-3-1}_{\mu} \to D^{\mathrm{SM}}_{\mu} =\partial_{\mu} -igT^3 W^3_{\mu} -i  \frac{ g t}{\sqrt{6+\beta^2 t^2}}\left( \beta T^8 + \sqrt{6} T^9 X\right) B_{\mu},	
\end{align*}
which results in the consequences that $g$ and $\frac{ g t}{\sqrt{6+\beta^2 t^2}}=gt_W$  are the gauge couplings of the SM, and the $U(1)_Y$ charge of the SM is  $Y/2=\beta T^8 +\mathbb{I}  X$.

Like the 331RN model, the 331ISS model includes two pairs of singly charged gauge bosons  with the following  physical states  $W^{\pm}$  and $Y^{\pm}$ and masses
\begin{align}
W^{\pm}_{\mu}&=\frac{W^1_{\mu}\mp i W^2_{\mu}}{\sqrt{2}},\; Y^{\pm}_{\mu}=\frac{W^6_{\mu}\pm i W^7_{\mu}}{\sqrt{2}},\; m_W^2=\frac{g^2}{4}\left(v_1^2+v_2^2\right),\;
m_Y^2=\frac{g^2}{4}\left(w^2+v_1^2\right). \label{singlyG}
\end{align}
The bosons $W^{\pm}$ are identified with the SM ones, leading to the consequence  that
\begin{align} \label{eq_SMvev}
v_1^2+v_2^2\equiv v^2=(246 \mathrm{GeV})^2.
\end{align}
The general Higgs potential relating with the 331RN  model will be applied in our work with $v_1\neq v_2$.
We will use the following parameters  for this general case.
\be\label{eq_tbeta}
	t_{\beta}\equiv \tan\beta=\frac{v_2}{v_1}, \quad v_1=vc_{\beta}, \quad v_2=vs_{\beta}.
\ee
The parameter $t_\beta$ plays a similar role  known in the well-known models with two Higgs doublet   and the minimal supersymmetric Standard Model.  This is  different from Ref.~\cite{Nguyen:2018rlb}, where $v_1=v_2$ was  assumed so that the Higgs potential given in Ref.~\cite{Hue:2015fbb}
  was used to find the exact physical state of the SM-like Higgs boson.  This simple condition was also used in previous discussions in 3-3-1 models addressed with anomalous magnetic dipole moments~\cite{deJesus:2020upp, deJesus:2020ngn}. As we will show below,  large $t_{\beta}\neq 1$ is one of the key condition for predicting large  $(g-2)_{\mu}$ consistent with experiments. The reason is that the physical states of the  charged Higgs bosons are determined analytically from this Higgs potential, and only these Higgs bosons contribute significantly to one-loop corrections to the $(g-2)_{\mu}$.

The Yukawa Lagrangian for generating lepton masses is:
\begin{align}
\label{eq_Lylepton}
\mathcal{L}^{\mathrm{Y}}_l =-h^e_{ab}\overline{\psi_{aL}}\rho e_{bR}+
h^{\nu}_{ab} \epsilon^{ijk} \overline{(\psi_{aL})_i}(\psi_{bL})^c_j\rho^*_k
- Y_{ab}\overline{\psi_{aL}}\,\chi X_{bR} -\frac{1}{2} (\mu_{X})^*_{ba}\overline{(X_{aR})^c}X_{bR}+ \mathrm{H.c.}.
\end{align}
Here  we assumed that the model under consideration respects a new lepton number symmetry $\mathcal{L}$   discussed  in Ref.  \cite{Chang:2006aa}  so that   the term   $\overline{\psi_{aL}}\,\eta X_{bR}$ is not allowed in the above Yukawa Lagrangian, while the soft-breaking term $(\mu_{X})^*_{ba}\overline{(X_{aR})^c}X_{bR}$ is allowed with small $(\mu_{X})_{ba}$. The new lepton number $\mathcal{L}$ called by generalized lepton number \cite{CarcamoHernandez:2017cwi} is defined as $L=\frac{4}{\sqrt{3}} T^8 + \mathcal{L} \mathbb{I}$, where $L$ is  the normal lepton number.  The specific assignment of $\mathcal{L}$ is  $\mathcal{L}(\rho)=-1/3$, $\mathcal{L}(\eta)=-2/3$, $\mathcal{L}(\chi)=4/3$, $\mathcal{L}(\psi_{aL})=1/3$, which guarantees the consistence for the well-known definition of $L$, namely $L(\ell)=1$ for $\ell=e_{aL,R}, \nu_{aL}$,  $L(\ell)=-1$ for $\ell=N_{aL},X_{aR}$, and $L(q)=0$ for  all SM quarks  \cite{Chang:2006aa}.

The first term in Lagrangian \eqref{eq_Lylepton} generates charged lepton masses $m_{e_a}\equiv \frac{h^e_{ab}v_1}{\sqrt{2}} \delta_{ab}$,
 i.e, the mass matrix of the charged leptons is assumed to be diagonal,  hence  the flavor states of the charged leptons are also the physical ones.
In the basis $\nu'_{L}=(\nu_{L}, N_{L}, (X_R)^c)^T$ and $(\nu'_{L})^c=((\nu_L)^c, (N_L)^c, X_R)^T$ of the neutral leptons,   Lagrangian \eqref{eq_Lylepton} gives a neutrino mass term corresponding to a  block form of the mass matrix~\cite{Nguyen:2018rlb}, namely
\begin{align}
-{\mathcal{L}}^{\nu}_{\mathrm{mass}}=\frac{1}{2}\overline{\nu'_L}M^{\nu\dagger }(\nu'_L)^c +\mathrm{H.c.}, \,\mathrm{ where }\quad M^{\nu\dagger}=\begin{pmatrix}
0	& m_D &0 \\
m_D^T	&0  & M_R \\
0& M_R^T& \mu_X^{\dagger}
\end{pmatrix},  \label{Lnu1}
\end{align}
where   $M_R$  is  a $3\times3$ matrix  $(M_R)_{ab}\equiv Y_{ab}\frac{w}{\sqrt{2}}$, $(m_D)_{ab}\equiv \sqrt{2}h^{\nu}_{ab}v_1$ with $a,b=1,2,3$. Neutrino subbases are denoted as  $\nu_{R}=((\nu_{1L})^c,(\nu_{2L})^c,(\nu_{3L})^c)^T$, $N_R=((N_{1L})^c,(N_{2L})^c,(N_{3L})^c)^T$, and  $X_L=((X_{1R})^c,(X_{2R})^c,(X_{3R})^c)^T$. The mass matrix $M_R$ does not appear in the 331RN.  The Dirac neutrino mass matrix $m_D$ must be antisymmetric.  The matrix  $\mu_{X}$ defined in Eq.~\eqref{eq_Lylepton}  is symmetric and it can be diagonalized by a transformation $U_X$:
\be\label{eq_muX}
	U_X^T\mu_{X}U_X=\mathrm{diag}\left(\mu_{X,1},\mu_{X,2},\mu_{X,3}\right).
\ee
The matrix $U_X$ will be absorbed by redefinition the  states $X_{a}$, therefore $\mu_X$ will be set as the diagonal matrix given in the right hand side of  Eq.~\eqref{eq_muX}.

The mass matrix  $M^{\nu}$ is   diagonalized by a $9\times9$ unitary matrix $U^{\nu}$,
\begin{align}
U^{\nu T}M^{\nu}U^{\nu}=\hat{M}^{\nu}=\mathrm{diag}(m_{n_1},m_{n_2},..., m_{n_{9}})=\mathrm{diag}(\hat{m}_{\nu}, \hat{M}_N), \label{diaMnu}
\end{align}
where $m_{n_i}$ ($i=1,2,...,9$) are  masses of the nine physical neutrino states $n_{iL}$. They consist of  three active neutrinos  $n_{aL}$ ($a=1,2,3$) corresponding to the mass  submatrix    $\hat{m}_{\nu}=\mathrm{diag}(m_{n_1},\;m_{n_2},\;m_{n_3})$, and the six extra neutrinos $n_{IL}$ ($I=4,5,..,9$)  with  $\hat{M}_N=\mathrm{diag}(m_{n_4},\;m_{n_5},...,\;m_{n_{9}})$.
The ISS mechanism leads to the following approximation solution of $U^{\nu}$,
\begin{align}
\label{eq_Unu}
U^{\nu}= \Omega \left(
\begin{array}{cc}
U_{\mathrm{PMNS}} & \mathbf{O} \\
\mathbf{O} & V \\
\end{array}
\right), \;\; \Omega=\exp\left(
\begin{array}{cc}
\mathbf{O} & R \\
-R^\dagger & \mathbf{O} \\
\end{array}
\right)=
\left(
\begin{array}{cc}
1-\frac{1}{2}RR^{\dagger} & R \\
-R^\dagger &  1-\frac{1}{2}R^{\dagger} R\\
\end{array}
\right)+ \mathcal{O}(R^3),
\end{align}
where
\begin{align}
R^* &\simeq \left(-m^*_DM^{-1}, \quad  m^*_D(M_R^\dagger)^{-1}\right), \quad  M\equiv M^*_R\mu_X^{-1}M_R^{\dagger}, \label{eq_R}\\
m^*_DM^{-1} m^\dagger_D&\simeq m_{\nu}\equiv U^*_{\mathrm{PMNS}}\hat{m}_{\nu}U^{\dagger}_{\mathrm{PMNS}},  \label{eq_mnu}\\
V^* \hat{M}_N V^{\dagger}& \simeq  M_N+ \frac{1}{2}R^TR^* M_N+ \frac{1}{2} M_NR^{\dagger} R,\; M_N\equiv \begin{pmatrix}
	0&M_R^*  \\
	M_R^{\dagger}& \mu_X
\end{pmatrix}
.  \label{eq_V}
 \end{align}
The relations between the flavor  and mass eigenstates are
\be
\nu'_L=U^{\nu} n_L, \quad  \mathrm{and} \; (\nu'_L)^c=U^{\nu*}  (n_L)^c, \label{Nutrans}
\ee
where $n_L\equiv(n_{1L},n_{2L},...,n_{9L})^T$ and $(n_L)^c\equiv((n_{1L})^c,(n_{2L})^c,...,(n_{9L})^c)^T$.
 The standard form of the lepton mixing matrix $U_{\mathrm{PMNS}}$ is the function   of three angles $\theta_{ij}$, one Dirac phase $\delta$ and two Majorana phases $\al_{1}$, and $\al_2$~\cite{Tanabashi:2018oca}, namely
\begin{align}
	U^{\mathrm{PDG}}_{\mathrm{PMNS}}
	&= \begin{pmatrix}
		1	& 0 &0  \\
		0	&c_{23}  &s_{23}  \\
		0&  	-s_{23}& c_{23}
	\end{pmatrix}\,\begin{pmatrix}
		c_{13}	& 0 &s_{13}e^{-i\delta}  \\
		0	&1  &0  \\
		-s_{13}e^{i\delta}&  0& c_{13}
	\end{pmatrix}\,\begin{pmatrix}
		c_{12}	& s_{12} &0  \\
		-s_{12}	&c_{12}  &0  \\
		0& 0 	&1
	\end{pmatrix} \mathrm{diag}\left(1, e^{i\alpha_{1}},\,e^{i\alpha_{2}}\right) \label{eq_UnuPDG}
	\crn&=U^0_{\mathrm{PMNS}} \;\mathrm{diag}\left(1, e^{i\alpha_{1}},\,e^{i\alpha_{2}}\right),
\end{align}
where $s_{ij}\equiv\sin\theta_{ij}$, $c_{ij}\equiv\cos\theta_{ij}=\sqrt{1-s^2_{ij}}$, $i,j=1,2,3$ ($i<j$), $0\le \theta_{ij}<90\; [\mathrm{Deg.}]$ and $0<\delta\le 720\;[\mathrm{Deg.}]$. The Majorana phases are chosen in the range $-180\le\alpha_i\le 180$ [Deg.]

 In this paper, we will work on the normal ordered scheme (NO) of the active neutrino masses, which allows  $\delta=\pi$ using in this work. The respective best fit and the confidence level of 3$\sigma$  of the neutrino oscillation experimental data is given  as~\cite{Zyla:2020zbs}
 \begin{align}
	\label{eq_d2mijNO}
	&s^2_{12}=0.32,\;  0.273\le s^2_{12}\le 0.379;
	\crn&s^2_{23}= 0.547,\;  0.445\le s^2_{23}\le 0.599;
	\crn&s^2_{13}= 0.0216 ,\;  0.0196\le s^2_{13}\le 0.0241;
	\crn&\delta= 218 \;[\mathrm{Deg}] ,\;157\;[\mathrm{Deg}] \le \delta\le 349\;[\mathrm{Deg}] ;
	\crn &\Delta m^2_{21}=7.55\times 10^{-5} [\mathrm{eV}^2], \quad   7.05\times 10^{-5} [\mathrm{eV}^2]\leq\Delta m^2_{21}\leq8.24\times 10^{-5} [\mathrm{eV}^2] ;
	\crn& \Delta m^2_{32}=2.424\times 10^{-3} [\mathrm{eV}^2], \quad 2.334\times 10^{-3} [\mathrm{eV}^2]\leq\Delta m^2_{32}\leq2.524\times 10^{-3} [\mathrm{eV}^2].
\end{align}
	%
	%
	%
The above $CP$ phase is consistent with  the updated one given in Ref.~\cite{Abe:2019vii}, where the allowed range corresponding to $3\sigma$ confidence level  are $-3.41\le \de \le -0.03$ ($164.6 \le \delta \le 358.3$ [Deg.]) for the NO scheme.
The lepton mixing matrix defined in Eq.~\eqref{eq_mnu}  relates with the experimental  parameters appearing in Eq.\eqref{eq_d2mijNO} are~\cite{Tanabashi:2018oca}
\begin{align}
	\label{eq_sij}
s^2_{12}&=\frac{|\left( U_{\mathrm{PMNS}}\right)_{12}|^2}{1-|\left( U_{\mathrm{PMNS}}\right)_{13}|^2}, \;
s^2_{13}=|\left( U_{\mathrm{PMNS}}\right)_{13}|^2, \;
s^2_{23}=\frac{|\left( U_{\mathrm{PMNS}}\right)_{23}|^2}{1-|\left( U_{\mathrm{PMNS}}\right)_{13}|^2}.
\end{align}
Additionally,  it is easily to derive that
\begin{align}
	\label{eq_delta}
	e^{i\delta}&= \frac{c_{23}(c^2_{12} +ys^2_{12})}{s_{13}s_{23}s_{12}c_{12}(1 -y)},\; y=\frac{\left( U_{\mathrm{PMNS}}\right)_{22}\left( U_{\mathrm{PMNS}}\right)_{11}}{\left( U_{\mathrm{PMNS}}\right)_{12}\left( U_{\mathrm{PMNS}}\right)_{21}},
	\crn e^{i\alpha_1}&= \frac{ \left(U_{\mathrm{PMNS}}\right)_{12}c_{12}}{ |\left(U_{\mathrm{PMNS}}\right)_{11}|s_{12}}, \;   e^{i\left( \alpha_2 -\delta\right)}= \frac{ \left(U_{\mathrm{PMNS}}\right)_{13}c_{13}c_{12}}{ |\left(U_{\mathrm{PMNS}}\right)_{11}|s_{13}}.
\end{align}
  The detailed calculation shown in Ref.~\cite{Nguyen:2018rlb}, using the ISS relations, 
yields
\be \label{eq_mD}
m_D= zc_{\beta}\times  \tilde{m}_D,\;  \tilde{m}_D=\begin{pmatrix}
	0&x_{12}  &x_{13}  \\
	-x_{12}& 0 &1  \\
	-x_{13}& -1 &0
\end{pmatrix},
\ee
where $z=\sqrt{2} v\,h^{\nu}_{23}$ is assumed to be positive and real,
\begin{align}
x^*_{12}&=\frac{(m_{\nu})_{11}(m_{\nu})_{23}-(m_{\nu})_{13}(m_{\nu})_{12}}{(m_{\nu})_{12}(m_{\nu})_{33}-(m_{\nu})_{13}(m_{\nu})_{23}},\;
x^*_{13}=\frac{(m_{\nu})_{11}(m_{\nu})_{33}-(m_{\nu})^2_{13}}{(m_{\nu})_{12}(m_{\nu})_{33}-(m_{\nu})_{13}(m_{\nu})_{23}}.
\label{eq_nxij}	
\end{align}
We note that  the lightest active neutrino mass  is zero at the tree level, but  can be nonzero when loop-corrections are included~\cite{Chang:2006aa}. Also, the quantum effects can be considered for the charged lepton masses, so that the  regions predicting  large $\De a_{\mu}$  may be larger~\cite{Yin:2021yqy, Baker:2021yli} than the ones discussed in this work.  The perturbative limit requires that $h^{\nu}_{23}< \sqrt{4\pi}$, leading to the following upper bound of $z$,
\be\label{eq_upperz}
z < 1233\; [\mathrm{GeV}].
\ee
The two formulas in Eq.~\eqref{eq_nxij} were found in the general symmetric from of $M^{-1}$, namely
they are be found by using Eq.~\eqref{eq_mnu} for off-diagonal entries of $m_{\nu}$ to determine $\left(M^{-1}\right)_{ij}$, then insert them into the diagonal ones. The off-diagonal
elements of $M^{-1}$ are determined as follows:
\begin{align}
	\label{eq_inverseM}
	\left(M^{-1}\right)_{12}&=\frac{1}{2} \left[ x^*_{13}\left(M^{-1}\right)_{11} - \frac{\left(M^{-1}\right)_{22}}{x^*_{13}} - \frac{\left(m_{\nu}\right)_{13} + x^*_{13}\left(m_{\nu}\right)_{23}}{x^*_{12}x^*_{13} z^2} \right],
	\crn
	\left(M^{-1}\right)_{13}&= \frac{1}{2} \left[ x^*_{12}\left(M^{-1}\right)_{11} + \frac{\left(M^{-1}\right)_{33}}{x^*_{12}} - \frac{\left(m_{\nu}\right)_{12} + x^*_{12}\left(m_{\nu}\right)_{23}}{x^*_{12}x^*_{13} z^2} \right],
	\crn
	\left(M^{-1}\right)_{23}&= \frac{1}{2} \left[  \frac{x^{*2}_{12}\left(M^{-1}\right)_{22} -x^{*2}_{13}\left(M^{-1}\right)_{33}}{x^*_{12}x^*_{13}} + \frac{ x^*_{13}\left(m_{\nu}\right)_{12} - x^*_{12}\left(m_{\nu}\right)_{13}}{x^*_{12}x^*_{13} z^2} \right].
\end{align}
Hence all elements of the matrix $M^{-1}$ depend on only three complex parameters $\left( M^{-1}\right)_{ii}$ with $i=1,2,3$. When identifying with  $M^{-1}=\left(M_R^{\dagger}\right)^{-1}\mu_X\left(M_R^{*}\right)^{-1}$ given in Eq.~\eqref{eq_R}, six parameters $\mu_{X,i}$ and $\left( M^{-1}\right)_{ii}$  are determined  as functions of elements of $M_R$. In this work, we will consider  all elements of $M_R$ are free parameters, namely
\be\label{key}
	\left(M_R\right)_{ij}= zc_{\beta}\times  	\left(\widetilde{M}_R\right)_{ij},\; \left(\widetilde{M}_R\right)_{ij}\equiv  k_{ij},
\ee
where all $k_{ij}$ are assumed to be real for simplicity.  The ISS relations are valid with at least some  $|k_{ij}|\gg1$ and det$M\neq 0$. In the numerical investigation,  $m_{\nu}$ is determined  from the $3\sigma$  neutrino oscillation data through Eq.~\eqref{eq_mnu}.  The Dirac matrix $m_D$ is then determined by Eq.~\eqref{eq_mD}.   The free parameters   $k_{ij}$ and $z$  are assumed to be real, and $z$ is positive.  The three elements of the  matrix $\mu_X$ are  determined as functions of  these free parameters. The respective  formulas are lengthy hence they are not written down explicitly here.
 In our work, we only consider the case max$|\mu_{X,i}|\ll z$ hence  all $(\mu_{X})_{i}$ gives suppressed mixing elements in the total lepton mixing matrix $U^{\nu}$.  This condition will always  be checked numerically to derive the final results.

 In the numerical investigation,   the free parameters $z$ and $k_{ij}$ will be scanned  in the valid ranges to construct the total neutrino mass matrix defined in Eq.~\eqref{Lnu1}. After that,   the mass eigenstates  and the  total mixing matrix are calculated  numerically with at least 30 digits of precision.   Using the relations listed in Eqs.~\eqref{eq_sij} and \eqref{eq_delta}, we  reproduce all of the oscillation parameters  $\De m^2_{ij}$ and $s^2_{ij}$  then force them satisfying the $3\sigma$ allowed data. This will help us to collect the allowed values of $z$ and $k_{ij}$  
in evaluating  the cLFV branching ratios and $(g-2)_{\mu}$ data. We emphasize that the regions of the parameter space in  our numerical investigation 
 are more  general  than those  mentioned in Refs.~\cite{Nguyen:2018rlb,Dinh:2020pqn}.

The Lagrangian for quark masses was discussed previously~\cite{Chang:2006aa}. Here, we just remind the reader that the Yukawa couplings of the top quark must satisfy the perturbative limit $h^u_{33}<\sqrt{4\pi}$, leading to a lower bound   $v_2 >\frac{\sqrt{2}m_t}{\sqrt{4\pi}}$. Combining this  with the relations in Eqs.~\eqref{eq_SMvev} and ~\eqref{eq_tbeta} gives a lower bound  $t_{\beta}\ge0.3$, which will be used in the numerical discussion.

\subsection{Higgs bosons}
The Higgs potential  used here respect the new lepton number defined in Ref.~\cite{Chang:2006aa}, namely
\begin{align}
\label{eq_Vh}
V_{h}&= \sum_{S=\eta, \rho,\chi} \left[ \mu_S^2 S^{\dagger}S +\lambda_S \left(S^{\dagger}S\right)^2 \right]  + \lambda_{12}(\eta^{\dagger}\eta)(\rho^{\dagger}\rho)
+\lambda_{13}(\eta^{\dagger}\eta)(\chi^{\dagger}\chi)
+\lambda_{23}(\rho^{\dagger}\rho)(\chi^{\dagger}\chi)  \crn
& +\tilde{\lambda}_{12} (\eta^{\dagger}\rho)(\rho^{\dagger}\eta)
+\tilde{\lambda}_{13} (\eta^{\dagger}\chi)(\chi^{\dagger}\eta)
+\tilde{\lambda}_{23} (\rho^{\dagger}\chi)(\chi^{\dagger}\rho) +\sqrt{2} \omega f\left(\epsilon_{ijk}\eta^i\rho^j\chi^k +\mathrm{h.c.} \right),
\end{align}
where $f$ is a dimensionless parameter, which  $f\omega$ is the same as that used in previous works.
The minimum conditions of the Higgs potential  as well as the identification of the SM-like Higgs were discussed in detailed previously~\cite{Ninh:2005su, Hue:2015fbb}. The model always contains a light $CP$ even neutral Higgs boson identified with the SM-like Higgs boson confirmed experimentally. This Higgs boson gives suppressed contributions to $(g-2)_{\mu}$  hence we will ignore it from now on.  The model contains two pairs of singly charged Higgs bosons $H^{\pm}_{1,2}$ and Goldstone bosons of the  gauge bosons $W^{\pm}$ and $Y^{\pm}$, which are denoted as $G^{\pm}_W$ and $G^{\pm}_Y$, respectively. The masses of all charged Higgs bosons are \cite{Buras:2012dp, Hue:2017lak,Ninh:2005su} $m^2_{H^{\pm}_1}=  \left( \frac{\tilde{\lambda}_{12}v^2}{2} +\frac{fw^2}{s_{\beta}c_{\beta}} \right)$,  $m^2_{H^{\pm}_2}= (v^2 c_{\beta}^2+w^2)\left(\frac{\tilde{\lambda}_{23}}{2}  +ft_{\beta}\right)  $,  and $m^2_{G^{\pm}_{W}}=m^2_{G^{\pm}_{Y}}=0$. The relations between the original and mass eigenstates of the charged Higgs bosons  are \cite{Ninh:2005su}
\bea
\left( \begin{array}{c}
\eta^\pm\\
\rho_1^\pm
\end{array} \right)=\left(
\begin{array}{cc}
-s_{\beta } & c_{\beta } \\
c_{\beta } & s_{\beta } \\
\end{array}
\right) \left(\begin{array}{c}
G_W^\pm
\\ H_1^\pm
\end{array} \right), \quad \left( \begin{array}{c}
\rho_2^\pm \\
\chi^\pm
\end{array} \right) = \left( \begin{array}{cc}
- s_\theta & c_\theta \\
c_\theta & s_\theta
\end{array} \right) \left(\begin{array}{c}
G_Y^\pm
\\ H_2^\pm
\end{array} \right),
\label{EchargedH}
\eea
where $t_{\theta}=v_1/w$.

The model contains five $CP$-odd neutral scalar components. Three of them are Goldstone bosons of  the neutral gauge bosons $Z,Z'$ and $X^0$. The two remaining are physical states with masses $
m^2_{a_1}=f \left( c_{\beta } s_{\beta } v^2 +\frac{ \omega ^2}{t_{\beta }} \right)+\frac{\bar{\lambda }_{13}}{2} \left(s_{\beta }{}^2 v^2+\omega ^2\right), \quad  m^2_{a_2}=f \left(\frac{\omega ^2}{c_{\beta }s_{\beta
}}+c_{\beta } s_{\beta } v^2\right).$
As a consequence, the parameter $f$ must be positive. In addition, $f$ may be small so that charged Higgs boson masses can be around 1 TeV.

%
\section{\label{sec_amuFormula}Analytic formulas for one loop contributions to $\Delta a_{e_a}$  and cLFV decays $e_b \rightarrow e_a\gamma$}
All detailed  steps for calculation to derive the  couplings that give large one-loop contributions  were presented in Ref.~\cite{Nguyen:2018rlb}. We just collect the final results related with this work.  The condition $m_{e_b}>m_{e_a}$ is always used to define the one loop  form factors $c^X_{(ab)R}$ and $c^X_{(ba)R}$ introduced in Ref.~\cite{Crivellin:2018qmi}, which are different from our notations  by a relative factor $m_{e_b}$.

The relevant Lagrangian of charged gauge bosons is
\begin{align}
	\mathcal{L}^{\ell n V}=\overline{\psi_{aL}}\gamma^{\mu}D_{\mu}\psi_{aL}
	&\supset \frac{g}{\sqrt{2}} \sum_{i=1}^9\sum_{a=1}^3\left[  U^{\nu*}_{ai} \overline{ n_{i}}\gamma^\mu P_L e_{a}W^{+}_{\mu}  + U^{\nu*}_{(a+3)i}\overline{n_{i} }\gamma^\mu  P_L e_{a}Y^{+}_{\mu} \right],\label{eq_llv1}
\end{align}
corresponding to the following one-loop form factors:
\begin{align}
	\label{eq_cabVR}
	c^{W}_{(ab)R}&= \frac{e g^2 }{32\pi^2 m_W^2} \sum_{i=1}^9 U^{\nu}_{ai}U^{\nu*}_{bi} F_{LVV}\left( \frac{m^2_{n_i}}{m^2_W}\right), \crn
		c^{W}_{(ba)R}&= \frac{e g^2 m_{e_a}}{32\pi^2 m_W^2m_{e_b}} \sum_{i=1}^9 U^{\nu}_{bi}U^{\nu*}_{ai} F_{LVV}\left( \frac{m^2_{n_i}}{m^2_W}\right), \crn
	c^{Y}_{(ab)R}&= \frac{e g^2 }{32\pi^2 m_W^2} \sum_{i=1}^9  U^{\nu}_{(a+3)i}U^{\nu*}_{(b+3)i}\frac{m^2_W}{m^2_Y}\times F_{LVV}\left( \frac{m^2_{n_i}}{m^2_Y}\right),\crn
		c^{Y}_{(ba)R}&= \frac{e g^2 m_{e_a}}{32\pi^2 m_W^2m_{e_b}} \sum_{i=1}^9  U^{\nu}_{(b+3)i}U^{\nu*}_{(a+3)i}\frac{m^2_W}{m^2_Y}\times F_{LVV}\left( \frac{m^2_{n_i}}{m^2_Y}\right),
	\end{align}
where
\be\label{eq_FFVV}
	F_{LVV}(x)= -\frac{10 -43 x +78 x^2 -49 x^3 +4 x^4 +18 x^3\ln(x)}{24(x-1)^4},
\ee
 $e=\sqrt{4\pi \alpha_{\mathrm{em}}}$ being the electromagnetic coupling constant, and $g=e/s_W$.

 Lagrangian of charged Higgs bosons is
 \begin{align}
 	\label{eq_LHee}
 	\mathcal{L}^{\ell n H}=- \frac{g}{\sqrt{2}m_W} \sum_{k=1}^2 \sum_{a=1}^3\sum_{i=1}^9 H_k^+\overline{n_i} \left(\lambda^{L,k}_{ai}P_L+\lambda^{R,k}_{ai}P_R\right)e_a   +H.c.,
 \end{align}
 where
 \begin{align}
 	\lambda^{R,1}_{ai}&=m_{e_a}U^{\nu*}_{ai}t_{\beta}, \quad 	
 	\lambda^{R,2}_{ai}=\frac{m_{e_a}c_{\theta}U^{\nu*}_{(a+3)i}}{c_{\beta}},
\crn\lambda^{L,1}_{ai}&= -t_{\beta}\sum_{c=1}^3(m_D^*)_{ac}U^{\nu}_{(c+3)i}= -s_{\beta}z\sum_{c=1}^3(\tilde{m}_D^*)_{ac}U^{\nu}_{(c+3)i},
\crn\lambda^{L,2}_{ai}&= \sum_{c=1}^3\frac{c_{\theta}}{c_{\beta}}\times \left[ (m_D^*)_{ac}U^{\nu}_{ci} + t^2_{\theta}(M_R^*)_{ac}U^{\nu}_{(c+6)i}\right]
 	\crn&= c_{\theta}z \sum_{c=1}^3 \left[ (\tilde{m}_D^*)_{ac}U^{\nu}_{ci} + t^2_{\theta}(\widetilde{M}_R^*)_{ac}U^{\nu}_{(c+6)i}\right] .
 	\label{eq_lambdaLR}
 \end{align}
  The one-loop form factors are:
 \begin{align}
 	\label{eq_cabHR}
 	c^{H,k}_{(ab)R}&= \frac{e g^2}{32\pi^2m^2_W m_{e_b}m^2_{H_k}} \sum_{i=1}^9\left[ \lambda^{L,k*}_{ai} \lambda^{R,k}_{bi} m_{n_i}F_{LHH}\left(\frac{m^2_{n_i}}{m^2_{H_k}} \right)
 	\right. \crn& \quad \left.
 	+  \left( m_{e_b}\lambda^{L,k*}_{ai} \lambda^{L,k}_{bi} +m_{e_a} \lambda^{R,k*}_{ai} \lambda^{R,k}_{bi}\right) \tilde{F}_{LHH}\left(\frac{m^2_{n_i}}{m^2_{H_k}} \right)\right], \crn
 		c^{H,k}_{(ba)R}&= \frac{e g^2}{32\pi^2m^2_W m_{e_b}m^2_{H_k}} \sum_{i=1}^9\left[ \lambda^{L,k*}_{bi} \lambda^{R,k}_{ai} m_{n_i}F_{LHH}\left(\frac{m^2_{n_i}}{m^2_{H_k}} \right)
 	\right. \crn& \quad \left.
 	+  \left( m_{e_a}\lambda^{L,k*}_{bi} \lambda^{L,k}_{ai} +m_{e_b} \lambda^{R,k*}_{bi} \lambda^{R,k}_{ai}\right) \tilde{F}_{LHH}\left(\frac{m^2_{n_i}}{m^2_{H_k}} \right)\right],
 \end{align}
 where  $b\ge a$, and
 \be  \label{eq_FLHH}
 	F_{LHH}(x)= -\frac{1 -x^2 + 2x\ln(x)}{4(x-1)^3}, \quad \widetilde{F}_{LHH}(x)= -\frac{-1 +6 x -3 x^2  -2 x^3  + 6 x^2\ln(x)}{24(x-1)^4}.
 \ee
The total one-loop contribution to the cLFV and $\De a^{331\mathrm{ISS}}_{\mu}$ is
  \begin{align} \label{eq_cabR}
 	c_{(ab)R}&=c^W_{(ab)R}+c^Y_{(ab)R}+ c^{H_1}_{(ab)R} +c^{H_2}_{(ab)R}, \crn
 	c_{(ba)R}&=\left\{ c_{(ab)R}\left[a\leftrightarrow b\right]\right\} \times \frac{m_{e_a}}{m_{e_b}}.
 \end{align}
The  one-loop contributions from charged gauge bosons  to the $a_{e_a}$ and the electric dipole moment $d_{e_a}$ of the  charged lepton $e_a$ are \cite{Crivellin:2018qmi}:
\begin{align}
\label{eq_amuWY}
a^V_{e_a}&= a^W_{e_a} +a^Y_{e_a}\equiv -\frac{4 m^2_{e_a}}{e} \left(\mathrm{Re}[c^{W}_{(aa)R}] +\mathrm{Re}[c^{Y}_{(aa)R}]\right),\crn
d^V_{e_a}&= d^W_{e_a} +d^Y_{e_a}\equiv -2m_{e_a} \left(\mathrm{Im}[c^{W}_{(aa)R}] +\mathrm{Im}[c^{Y}_{(aa)R}]\right),
\end{align}
The  one-loop contribution to $a_{e_a}$ and $d_{e_a}$ caused by charged Higgs bosons is~\cite{Crivellin:2018qmi}:
\begin{align} \label{eq_amuHee}
a^{H}_{e_a}&=\sum_{k=1}^2a^{H,k}_{e_a}, \; a^{H,k}_{e_a}\equiv -\frac{4 m^2_{e_a}}{e} \mathrm{Re}[c^{H,k}_{(aa)R}] ,\crn
d^{H}_{e_a}&=\sum_{k=1}^2d^{H,k}_{e_a}, \; d^{H,k}_{e_a}\equiv -2 m_{e_a} \mathrm{Im}[c^{H,k}_{(aa)R}].
 \end{align}
The quantity $\Delta d_{e_a}= d^V_{e_a}+d^H_{e_a}$ is the new one loop contributions predicted to the  electric dipole moment  of the charged leptons.  It equals to  zero when our investigation is limited in the case  of the Dirac phase $\delta=\pi$. This zero value of $d_{\mu}$ satisfies the current experimental constraint~\cite{Muong-2:2008ebm} hence we will  not consider from now on.

We remind the reader that one loop contributions from neutral Higgs bosons are very suppressed hence they are ignored  here. The reason is that the 331ISS model  has no new charged leptons, hence the one-loop contributions of any neutral Higgs bosons $H^0$  to $c_{(ab)R}$ must arise only from the  couplings $ H^0\bar{e}_ae _a$ derived from the first term of the Yukawa Laragian~\eqref{eq_Lylepton}. These couplings have the same Yukawa couplings with the SM-like  $h\sim \mathrm{Re}[\rho^0]/\sqrt{2}$, but different mixing factors $|c_{H^0}|\le 1$ telling the contributions of $\rho^0$ to the physical state $H^0$. Hence these  contributions to $a_{\mu}$ have the same form with the one from the SM-like Higgs boson having mass $m_h\simeq 125$ GeV $\gg m_{\mu}$, $a^{h}_{\mu}\simeq \frac{\sqrt{2} G_{\mu} m^2_{\mu}}{4\pi^2} \times \frac{m^2_{\mu}}{m_h^2} \ln\frac{m_h^2}{m^2_{\mu}}\le  \mathcal{O}(10^{-14})$  \cite{Jegerlehner:2009ry}. Also,  the heavy neutral Higgs  will give  suppressed one-loop contributions to $\Delta a_{\mu}$. The deviation of $a_{\mu}$ between predictions by  the two models 331ISS and SM are
\begin{align}
\label{eq_defDamu331}
\Delta a^{\mathrm{331ISS}}_{e_a}&\equiv \Delta a_{e_a}= \Delta a^{W}_{e_a} + a^{Y}_{e_a} + a^{H,1}_{e_a}  +a^{H,2}_{e_a}, \quad  \Delta a^{W}_{e_a}= a^{W}_{e_a} - a^{\mathrm{SM},W}_{e_a},
\end{align}
where $a^{\mathrm{SM},W}_{\mu}=3.887\times 10^{-9}$~\cite{Jegerlehner:2009ry} is the SM prediction for the one-loop contribution from $W$ boson $c^{W,\mathrm{SM}}_{(22)R}$.  In this work, $\De a^{\mathrm{331ISS}}_{\mu}=\De a_{\mu}$ will be considered as new physics predicted by the 331ISS and will be used to compare with the experimental data in the following numerical investigation.  We note that the discrepancy of $a_e$ between experiments and SM  is  about 2.5 standard deviation \cite{Parker:2018vye, Aoyama:2012wj, Aoyama:2012wk, Laporta:2017okg, Terazawa:2018pdc, Volkov:2019phy}. In this work we  will only pay attention to the $\De a_{\mu}$ which is the very interesting result of 4.2 standard deviation and may be a clear signal of   new physics in the near future.

Based on Ref.~\cite{Crivellin:2018qmi}, the branching ratios   of the cLFV  processes  are
\be\label{eq_Gaebaga}
\mathrm{Br}(e_b\rightarrow e_a\gamma)\simeq \frac{48\pi^2}{ G_F^2} \left( \left|c_{(ab)R}\right|^2 +\left|c_{(ba)R}\right|^2\right) \mathrm{Br}(e_b\rightarrow e_a\overline{\nu_a}\nu_b),
\ee
where $G_F=g^2/(4\sqrt{2}m_W^2)$.
 This result is consistent with the  formulas given used in Refs.~\cite{Hue:2017lak, Nguyen:2018rlb} for 3-3-1 models.

 It is noted that for the gauge boson contributions, we have  $|c^V_{(ba)R}|/|c^V_{(ab)R}|=m_{e_a}/m_{e_b}\ll1$ for $m_{e_b}>m_{e_a}$.  Similarly, we can estimate that $|c^{H,k}_{(ba)R}|/|c^{H,k}_{(ab)R}|\ll1$ for every particular contribution.   Anyways,  in the general case we cannot ignore $c^{X}_{(ba)R}$ because of the situation that  when  contributions to $c_{(ab)R}$ have the same order but some of them have opposite signs. Then the very destructive correlations among particular Higgs  contributions  in the $c_{(ab)R}$ will result in the same order of both $|c_{(ab)R}|$ and $|c_{(ba)R}|$.
  This will happen in the 331ISS model when  $\De a^{331\mathrm{ISS}}_{\mu}=\mathcal{O}(10^{-9})$ corresponding to the order of the experimental data    and Br$(\mu \rightarrow e\gamma)<4.2\times 10^{-13}$ require both  conditions of $ \mathcal{O}(10^{-9}) \; [\mathrm{GeV}^{-2}]\le|c_{(22)R}|\le \mathcal{O}(10^{-8}) \; [\mathrm{GeV}^{-2}]$ and $|c_{(21)R}|\le \mathcal{O}(10^{-13})\;[\mathrm{GeV}^{-2}]$, respectively.  As a result, we can estimate  that the  one-loop contributions from two charged Higgs bosons to Br($\mu\rightarrow e\gamma$) are strongly destructive, i.e.  $c^{H_1}_{(12)}\simeq- c^{H_2}_{(12)}$. Simultaneously,   $|c^{H_k}_{(12)}| \sim |c^{H_k}_{(22)}|$, therefore  the charged Higgs  contributions to $\De a_{\mu}$ must  be  constructive  and satisfy  $|c^{H_1}_{(22)}|\sim |c^{H_2}_{(22)}| \sim \mathcal{O} (10^{-9})-\mathcal{O} (10^{-8})\; [\mathrm{GeV}^{-2}]$, or they can be destructive but $|c^{H_i}_{(22)}|\gg |c^{H_j}_{(22)}| $ with $i\neq j$. These important properties of charged Higgs boson  contributions will be the key point in our  numerical investigation to collect data points  satisfying the  large values of  $\Delta a_{\mu}\ge 10^{-9}$ before considering any cLFV decay constraints.
  The gauge contributions  are suppressed hence we do not discuss qualitatively here, but they are also included in the numerical investigation.  We just pay attention to the two key one-loop charged Higgs boson contributions which will affect two other cLFV decays   $\tau \rightarrow e\gamma, \mu \gamma$.

 The experimental constraints of the form factors $c_{(ab)R}$  are listed in Table~\ref{table_cabRcontraint}, where the allowed values of $\Delta a_{\mu}$ are chosen in the range of $1\sigma$ confidence level given in Eq.~\eqref{eq_damu}.
\begin{table}[ht]
	\centering
	\begin{tabular}{cc}
		\hline
$192 <\Delta a_{\mu}\times 10^{11}<310,$ 	& $-4.8\times 10^{-8} \;  [\mathrm{GeV}^{-2}] <c_{(22)R}< -3.99 \times 10^{-8}\; [\mathrm{GeV}^{-2}]$\\
Br$(\mu \rightarrow e\gamma)$ & $|c_{(21)R}|,\; |c_{(12)R}|<3.47\times 10^{-13}\; [\mathrm{GeV}^{-2}]$\\
Br$(\tau \rightarrow e\gamma)$& $|c_{(31)R}|,\; |c_{(13)R}|<2.31\times 10^{-10}\; [\mathrm{GeV}^{-2}]$\\
Br$(\tau \rightarrow \mu \gamma)$ & $|c_{(32)R}|,\; |c_{(23)R}|< 2.63\times 10^{-10}\; [\mathrm{GeV}^{-2}]$\\
		\hline
	\end{tabular}
	\caption{Constraints of  $c_{(ab)R}\; [\mathrm{GeV}^{-2}]$ from experimental data. The allowed values of $\Delta a_{\mu}$ satisfying a confidence level of $1\sigma$  from the experimental data given in Eq.~\eqref{eq_damu}.}\label{table_cabRcontraint}
\end{table}
We derive that the allowed regions of the parameter space have the following properties:
\begin{align}
\label{eq_cabRcontraints}
\left| \frac{c_{(12)R}}{c_{(22)R}}\right|,  \left| \frac{c_{(21)R}}{c_{(22)R}}\right|\le \mathcal{O}(10^{-5});  \;
\left| \frac{c_{(13)R}}{c_{(22)R}}\right|, \;  \left| \frac{c_{(31)R}}{c_{(22)R}}\right|, \; \left| \frac{c_{(23)R}}{c_{(22)R}}\right|, \; \left| \frac{c_{(32)R}}{c_{(22)R}}\right|  \le \mathcal{O}(10^{-2}).
\end{align}
Normally,  our numerical scan gives  a relation that  $|c^{H_k}_{(22)R}|/ |c^{H_k}_{(ab)R}|\le \mathcal{O}(10)$ with $a\neq b$.
As a result, the huge destructive correlation between charged Higgs contributions to guarantee simultaneously the experimental constraints of Br$(\mu\rightarrow e\gamma)$ and $\De a_{\mu}$.  Also,  the two cLFV decays of $\tau \rightarrow e\gamma,\mu\gamma$  also need smaller but still large destructive charged Higgs contributions to satisfy the  upper experimental bounds because  some of these particular contributions often satisfy $|c^{H_k}_{(13)R}|/|c^{H_k}_{(22)R}|,|c^{H_k}_{(23)R}|/|c^{H_k}_{(22)R}|\ge 0.1$. While $|c^{H_k}_{(31)R}|,|c^{H_k}_{(32)R}|\ll 10^{-10}\; [\mathrm{GeV}^{-2}]$, 
consequently they are subdominant to the cLFV decays where their branching ratios are close to the upper experimental constraints. The mentioned properties are very important for us to point out the validation of the allowed regions.

For convenience in estimating qualitatively the above properties, we define  new important quantities determining the correlations between two charged Higgs contributions in a physical process as follows:
\begin{align}
	R^X_{ab}&\equiv \left|\frac{\mathrm{Re}[c^{X}_{(ab)R}] }{\mathrm{Re}[c_{(ab)R}]} \right|; \; a,b=1,2,3; \; X=W,Y,H^{\pm}_1,H^{\pm}_2, 	\label{eq_RabX} \\
	R^{-}_{ab}&\equiv \left|\frac{\mathrm{Re}[c^{H_1}_{(ab)R}+c^{H_2}_{(ab)R}] }{\mathrm{Re}[c_{(ab)R}]}\right|.  \label{eq_Rmab}
\end{align}
The first ratio $R^X_{ab}$ shows the relative contribution from the particle $X$ in the loop to the total contribution. The second one shows the relative  contributions of both singly charged Higgs  bosons.    In the 331ISS model, we will see that the relations $R^W_{ab},R^Y_{ab}\ll R^{H_k}_{ab}$ often happens. The interesting possibility we would like to discuss is that large contributions of $H_k$ for large $\Delta a_{\mu}$ with $|R^{H_k}_{22}|\sim 10^{-1}$, while the huge destructive correlations of these two Higgs bosons $\mathrm{Re}[c^{H_1}_{(ab)R}/c^{H_2}_{(ab)R}]\simeq -1$ will allow small cLFV constraints. Quantitatively, we estimate that  $R^{H_k}_{ab}\gg 1$ and $R^-_{ab}\simeq 1$, with  $a\neq  b$.   The details of numerical investigation will be shown  below.

\section{ \label{sec_HiggSinglet} Additional  singly charged Higgs boson  for an explanation of $(g-2)_{\mu}$ data at 1 $\sigma$ deviation}

The appearance of the gauge singlet $X_R$ leads to a possibility that,  a new singly charged Higgs bosons $h_3^{\pm}\sim (1,1,\pm1)$   can be included in the 331ISS model so that they can  give one-loop contributions to both $\De a_{\mu}$ and cLFV amplitudes through the following Yukawa interactions:
\be\label{eq_Yukawah3}
	\mathcal{L}^Y_{h_3}=-Y^{3}_{ab} \overline{(X_{aR})^c}e_{bR}h^{+}_3 +\mathrm{H.c.}=-Y^{3}_{ab} U^{\nu*}_{(a+6)i}\overline{(n_{i})}P_Re_{b}h^{+}_3 +\mathrm{H.c.}.
\ee
The new contributions to the cLFV decays and $\De a^{331\mathrm{ISS}}_{\mu}$ is
\begin{align}
	\label{eq_cab3R}
	c^{h_3}_{(ab)R}&= \frac{em_{e_a}}{16\pi^2 m_{e_b}m^2_{h_3}} \sum_{i=1}^9 \sum_{c=1}^3 Y^{3}_{ca}Y^{3*}_{cb} U^{\nu*}_{(a+6)i} U^{\nu}_{(b+6)i}\tilde{F}_{LHH}\left(\frac{m^2_{n_i}}{m^2_{h_3}} \right)
	, \crn
	c^{h_3}_{(ba)R}&= \frac{e}{16\pi^2 m^2_{h_3}} \sum_{i=1}^9 \sum_{c=1}^3 Y^{3*}_{ca}Y^{3}_{cb} U^{\nu}_{(a+6)i} U^{\nu*}_{(b+6)i}\tilde{F}_{LHH}\left(\frac{m^2_{n_i}}{m^2_{h_3}} \right).
\end{align}
Although the contributions of these singly charged Higgs bosons  to $\De a_{\mu}$ are normally small and negative,  the contributions to the cLFV amplitudes  may be significantly large. 
Consequently, they can affect destructively the total cLFV decay amplitudes. These properties will keep   $\De a^{331\mathrm{ISS}}_{\mu}$ reaching  the experimental constraint given in Eq.~\eqref{eq_damu}, while keeping all other cLFV branching ratios well below the experimental constraints.
In this work,  we consider the simplest case that $h^\pm_3$  does not mix with the other singly charged Higgs bosons in the 331RN, and the mass is another free parameter. All of these properties can be derived easily from the total Higgs potential, hence it will be ignored in this work.

\section{\label{sec_numerical} Numerical discussion }
\subsection{Without contributions from additional singly charged Higgs bosons $h^\pm_3$}
The numerical experimental parameters are taken from Ref.~\cite{Zyla:2020zbs}:
\begin{align}
\label{eq_ex}
G_F&= 1.663787\times 10^{-5}\; \mathrm{GeV}^{-2},\; g=0.652,\; \alpha_e=\frac{1}{137}= \frac{e^2}{4\pi} ,\; s^2_{W}=0.231,\crn
m_e&=5\times 10^{-4} \;\mathrm{GeV},\; m_{\mu}=0.105 \;\mathrm{GeV},\; m_{\tau}=1.776\;\mathrm{GeV}, \; m_W=80.385 \; \mathrm{GeV},\crn
\mathrm{Br}(\mu \rightarrow e \overline{\nu_e}\nu_{\mu})&\simeq 1., \; \mathrm{Br}(\tau \rightarrow e \overline{\nu_e}\nu_{\tau})\simeq 0.1782, \; \mathrm{Br}(\tau \rightarrow \mu \overline{\nu_{\mu}}\nu_{\tau})\simeq 0.1739.
\end{align}
Before  discussing on the allowed regions that satisfy all  experimental constraints of cLFV decays $e_b\rightarrow e_a\gamma$ as well as $(g-2)_{\mu}$ data, we give some important crude estimation on the allowed regions of  parameter space constrained by both large    $\De a_{\mu}^{331\mathrm{ISS}}\ge \mathcal{O}( 10^{-9})$ and small Br$(e_b \rightarrow e_a\gamma)$.  The way to derive the total mass matrix to calculate numerically the masses and total neutrino mixing matrix $U^{\nu}$ were presented in the previous section.  We have checked that the input changes of $\De m^2_{ij}$ and $s^2_{ij}$ in the allowed ranges given in Eq.~\eqref{eq_d2mijNO} do  not change significantly the final results, so we will fix these quantities at their best-fit points. An exception that the Dirac phase $\delta=180$ [Deg.] is considered so that the imagine parts  of $c_{(ab)R}$ are zeros, leading to a simple case of destruction among the one-loop contributions from  charged Higgs bosons.

In the numerical scan, the  points  in the allowed regions also satisfy simultaneously  the following conditions:
\ben
	\item The condition    $\mathrm{Re}[c^{H_1}_{(ab)R}] /\mathrm{Re}[c^{H_2}_{(ab)R}]<0$ with $a\neq b$, will give a possibility that
	 $\mathrm{Re}[c^{H_1}_{(21)R}] +\mathrm{Re}[c^{H_2}_{(21)R}]\sim 0$,  which will result in  valid regions of the parameter space
	 in which  two charged Higgs bosons  contributions can cancel each others. Therefore, these regions will contain points  which give the very suppressed  total contributions to guarantee  the small  Br$(e_b \rightarrow e_a\gamma)$. We will use this  condition in our numerical investigation.
	
	\item A crude numerical scan shows that the condition $\mathrm{Re}[c^{H_1}_{(22)R}]/\mathrm{Re}[c^{H_2}_{(22)R}]>0$  so that the two charged Higgs bosons contributions to $\De a^{331\mathrm{ISS}}_{\mu}$ always have the same sign, i.e., they give constructive contributions. Therefore  the values of $\De a^{331\mathrm{ISS}}_{\mu}$ are remained  in the original orders of $\mathcal{O}(10^{-9})$.  Another case giving large $\De a_{\mu}$ is that   $|\mathrm{Re}[c^{H_i}_{(22)R}]|\ll |\mathrm{Re}[c^{H_j}_{(22)R}]|$ with $i\ne j$ when they have opposite signs.
\een

First,  we consider the simplest cases of all zero values of off-diagonal elements $k_{ij} = 0$ with $i\neq j$. The numerical investigation shows that we cannot obtain any allowed points satisfying simultaneously both experimental data of cLFV constraints  and $\De a_{\mu}$. The reason is  that there always exists a strict relation that $\mathrm{Re}[c^{H_1}_{(22)R}]/\mathrm{Re}[c^{H_2}_{(22)R}]$ and $\mathrm{Re}[c^{H_1}_{(21)R}]/\mathrm{Re}[c^{H_2}_{(21)R}]$ are always negative 
leading to small Br$(\mu\rightarrow e\gamma)$. As a 
consequence, charged Higgs contributions to $\De a^{331\mathrm{ISS}}_{\mu}$ are always destructive.
Hence, the derived values are smaller than the experimental data. A requirement of Br$(\mu\rightarrow  e\gamma)\le \mathcal{O}(10^{-8})$ gives largest values of $\De a^{331\mathrm{ISS}}_{\mu}<10^{-9}$.

From a crude numerical scan,   we  can  find the allowed regions of the parameter space  satisfying  both conditions that Br$(\mu \rightarrow e\gamma)< 4.2\times 10^{-13}$ and  large  $\Delta a_{\mu}^{331\mathrm{ISS}}\ge \mathcal{O}(10^{-9})$.  These allowed regions will be used to collect the allowed points satisfying the remaining cLFV constraints. The following  ranges  of the parameter space  will be chosen as the  necessary conditions of free parameters when  scanning  to collect allowed points:
 \begin{align}
 	\label{eq_scanNO1}
 	& t_{\beta}\in[0.3, 60], \;
 0.6\;\mathrm{[TeV]} \le m_{H_1},\; m_{H_2} \le 3 \;\mathrm{[TeV]},\crn
 	& \; |k_{ij}|\times zc_{\beta}< \sqrt{4\pi} w=5.3  \;\mathrm{[TeV]}, \;
 	10\;\mathrm{[GeV]} \le z \le 1223 \;\mathrm{[TeV]}.
 \end{align}
Numerical values of $k_{ij}$ will be chosen so that  they give  active neutrino masses and $U_{\mathrm{PMNS}}$  consistent with neutrino oscillation data.   The value of $5.3$ TeV is fixed from the  lower bound $w$ obtained from the experimental data of the heavy $Z'$ boson mass $m_{Z'}$. But it can be relaxed with larger $w$ without any changes of final conclusions in this work.

\begin{figure}[ht]
	\centering
	\begin{tabular}{cc}
		\includegraphics[width=7.5cm]{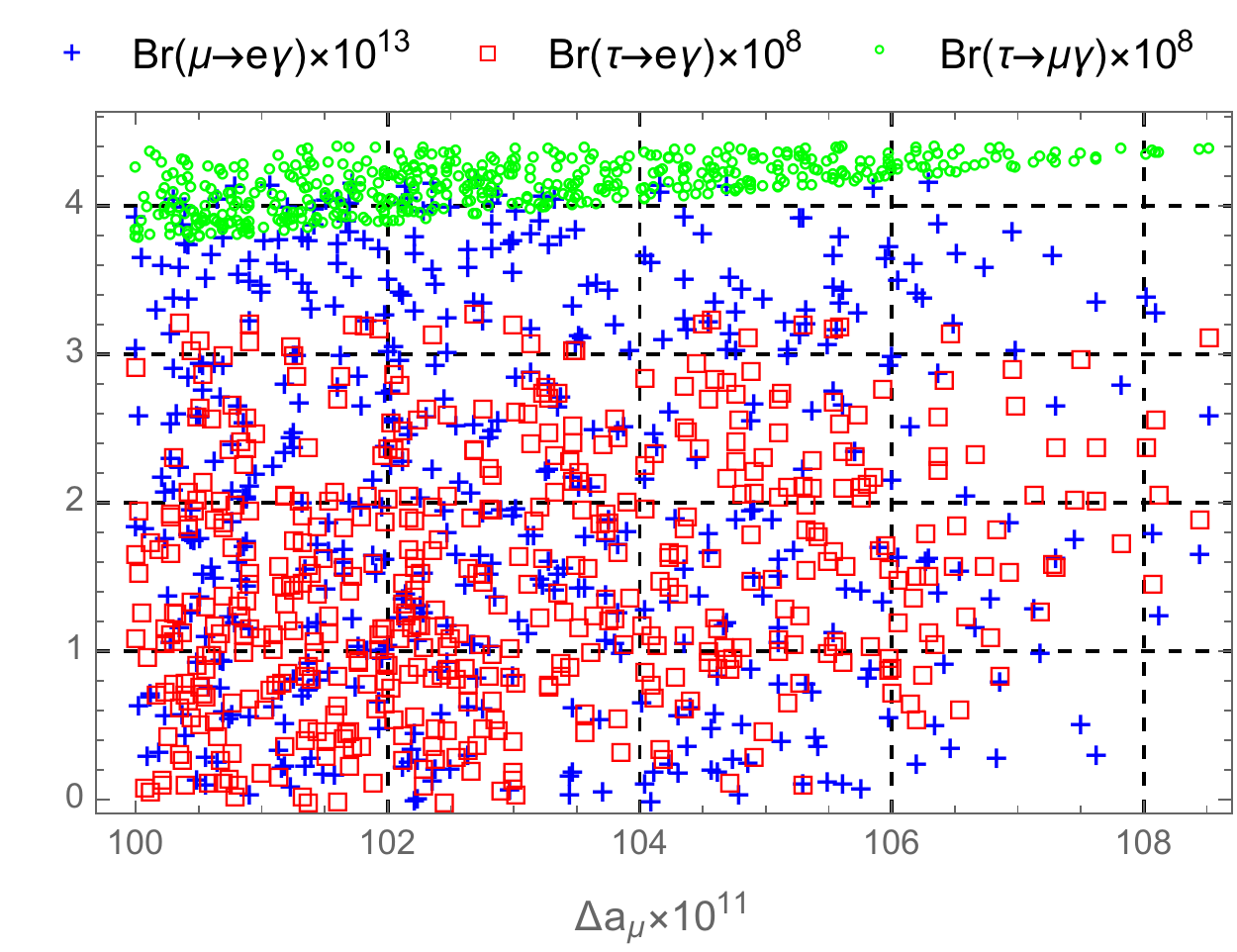}	&	
		\includegraphics[width=7.5cm]{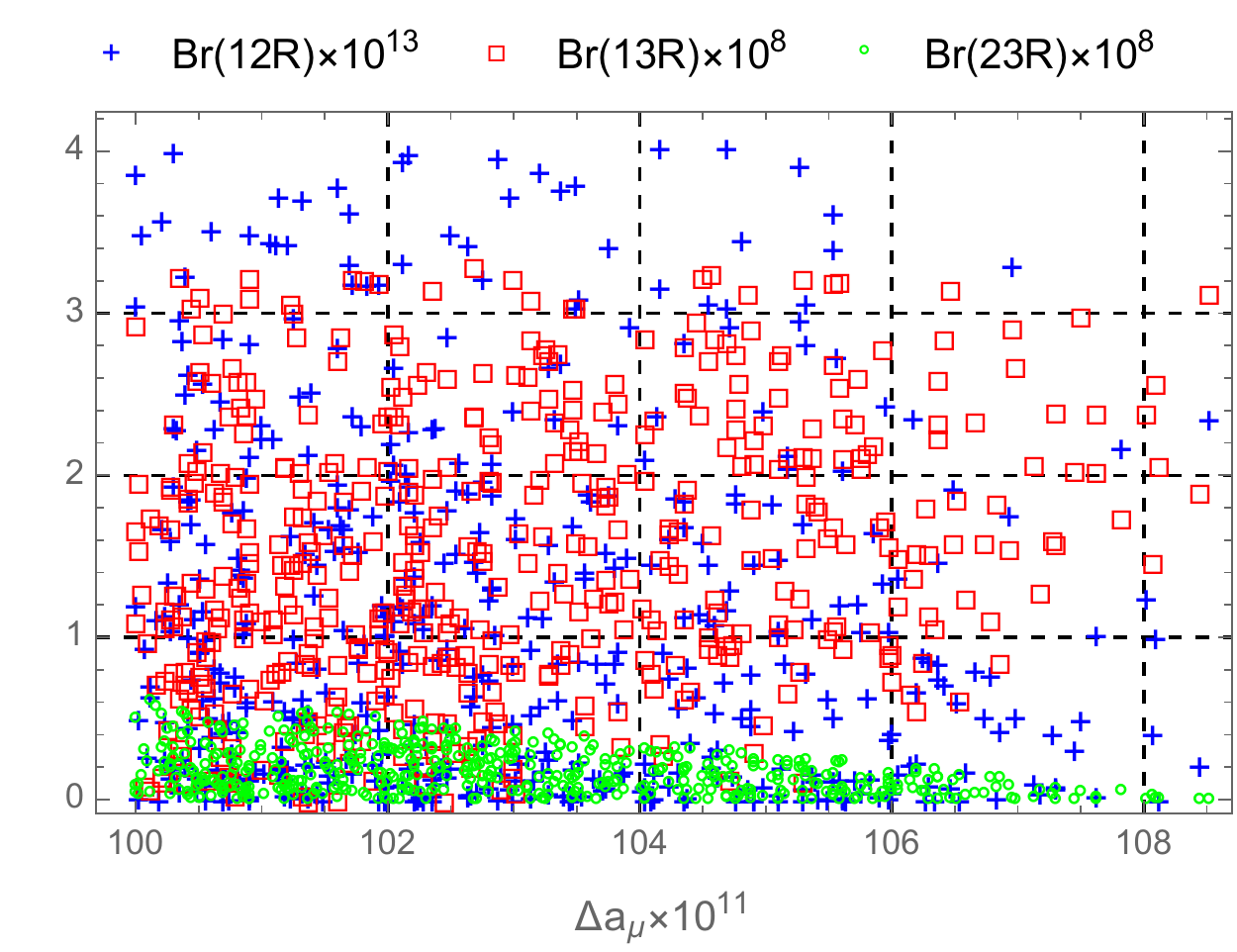}\\
		\end{tabular}
	\caption{  The left panel shows  $\Delta a_{\mu}$ vs. Br$(e_b\rightarrow e_a\gamma)\sim \left( |c_{(ab)R}|^2 +|c_{(ba)R}|^2\right)$  in the  free parameter ranges  given in Table~\ref{t_range1free}. The right panel shows  Br$(abR)\sim |c_{(ab)R}|^2$ with $a<b$. }\label{fig_amu-cLFV}
\end{figure}

Without contributions of the additional singly charged Higgs boson, our numerical investigation shows that  the largest values of $\De a_{\mu}$  satisfying  all cLFV constraints is $\De a_{\mu}\le 108.5 \times 10^{-11}$,  see an illustration  shown in Fig.~\ref{fig_amu-cLFV}.
The corresponding ranges of the free parameters are shown in Table~\ref{t_range1free},
\begin{table}[ht]
	\begin{tabular}{cccccccccccccc}
		\hline
	Notation	&$k_{11}$ &$ k_{22}$ & $k_{33}$ & $k_{12}$ & $k_{13}$ & $k_{23}$ & $k_{21}$ & $k_{31}$ & $k_{32}$ & $t_{\beta }$ & $z $ [GeV] & $m_{H_1}$[GeV] & $m_{H_2}$[GeV]\\	
		\hline 	
	Min	& -3.99 & -50.2 & 509. & -29.9 & 15.4 & -80.4 & 121. & 21.2 & 29.4 & 29.0 & 885. &705 & 769  \\
	Max& 2.47 & -35.1 & 528. & -20.6 & 24.9 & -66.4 & 135. & 36.8 & 45.9 & 40.0 & 1150 & 893 & 962 \\
	Example&   -3.26 & -49.7 & 509. & -28.6 & 23.0 & -77.8 & 124. & 25.3 & 37.1 & 36.9 & 969. & 754 & 825   \\
	\hline
	\end{tabular}
	\caption{ Numerical values of free parameters  for large $\Delta a^{331\mathrm{ISS}}_{\mu}\ge  10^{-9}$ satisfying all experimental  constraints of the cLFV decays $e_{b}\rightarrow e_a\gamma$.  \label{t_range1free}}
\end{table}
where the right panel shows the  only contributions from $c_{(ab)R}$ ($a<b$) to the decay rates, namely
$$ \mathrm{Br}(abR)=\frac{48\pi^2}{ G_F^2} \left|c_{(ab)R}\right|^2 \mathrm{Br}(e_b\rightarrow e_a\overline{\nu_a}\nu_b).$$
Here the two first lines show  the respective minimum and maximum values of the free parameters. The third line shows a particular example  of the set of the parameters  giving  large $\De a_{\mu}\simeq 108.1 \times 10^{-11}$.  The other quantities are shown in Table~\ref{t_largeDeamu}, which will be discussed more later.  In the left panel of Fig.~\ref{fig_amu-cLFV}, only Br$(\tau \rightarrow \mu \gamma)$ always enhances with increasing $\De a_{\mu}$.   The upper constraint Br$(\tau \rightarrow \mu \gamma)<4.4\times 10^{-8}$ gives the largest value of $\De a_{\mu}\simeq 108.5 \times 10^{-11}$.
From the right panel of Fig.~\ref{fig_amu-cLFV}, we see that $|c_{(23)R}|<|c_{(32)R}|$ in the region predicting large $\Delta a_{\mu}$, because the contribution from  $|c_{(23)R}|$  to  Br$(\tau \rightarrow \mu \gamma)$  denoted as  Br$(23R)$ is small,  namely  Br$(23R)\le 0.2\times 10^{-8}$ with  $\Delta a_{\mu}\ge 108\times 10^{-11}$. This is in contrast to  other normal  cases, as we will discuss based on the Table~\ref{t_largeDeamu}.

 Table~\ref{t_largeDeamu} illustrates   particular values of $c_{(ab)R}$ and large $\De a_{\mu}^{331\mathrm{ISS}}\simeq 108.1\times 10^{-11}$, corresponding to a set of free parameters given in the third line of Table~\ref{t_range1free}.
\begin{table}[ht]
	\begin{tabular}{ccccccc}
		\hline
Notations	&	$c^W_{(ab)R}-c^{W,\mathrm{SM}}_{(ab)R}$ &$c^Y_{(ab)R}$ & $c^{H_1}_{(ab)R}$ & $c^{H_2}_{(ab)R}$ & $c_{(ab)R}$ & Process \\
\hline
 $\Delta a_{\mu}:\;c_{(22)R}\times 10^{10}$& 5.22 & -0.499 & -82.07 & 3.11 & -74.24 & $ \Delta a_{\mu} =10.81 \times 10^{-10}$  \\
$\mu\rightarrow e\gamma:\;c_{(12)R}\times 10^{13}$&   422.13 & 29.645 & -29086. & 28636. & 1.6960 & $\; \mathrm{Br}(12R) = 1.002 \times 10^{-13}$ \\
$\mu\rightarrow e\gamma:\; c_{(21)R}\times 10^{13}$&  2.010 & 0.1412 & -138.5 & 138.9 & 2.568 &  $\;  \mathrm{Br}(21R) = 2.296 \times 10^{-13}$  \\
$\tau\rightarrow e\gamma:\; c_{(13)R}\times 10^{10}$&-0.031 & 0.01941 & 13.60 & -15.63 & -2.039 & $\;  \mathrm{Br}(13R) =  257.9 \times 10^{-10}$    \\
$\tau\rightarrow e\gamma:\; c_{(31)R}\times 10^{10}$& $\simeq0$& $\simeq0$& 0.004 & 0.031 & 0.035 & $\;  \mathrm{Br}(31R) = 0.076 \times 10^{-10}$  \\
$\tau\rightarrow \mu \gamma:\; c_{(23)R}\times 10^{10}$& -0.02505 & -0.03235 & -0.3305 & 0.5170 & 0.1291 & $\;  \mathrm{Br}(23R) =  1.009  \times 10^{-10}$  \\
$\tau\rightarrow \mu \gamma:\;c_{(32)R}\times 10^{10}$&  -0.001481 & -0.001913 & -0.01954 & -2.656 & -2.679 & $\; \mathrm{Br}(32R) =  434.7 \times 10^{-10}$  \\
\hline
	\end{tabular}
	\caption{ Particular contributions $c^X_{(ab)R}\;  [\mathrm{GeV}^{-2}]$ to the $\Delta a_{\mu}$ and Br$(e_b\rightarrow e_a\gamma)$  with the free parameters given in the third line of Table~\ref{t_range1free}. The last column shows values of $\Delta a_{\mu}$ and  Br$(e_b\rightarrow e_a\gamma)$. \label{t_largeDeamu}}
\end{table}
The numerical results given in Table~\ref{t_largeDeamu} show that  the experimental constraint from Br$(\tau\rightarrow \mu\gamma)<4.4\times 10^{-8}$  does not allow large  $\Delta a_{\mu}^{331\mathrm{ISS}}>108.5\times 10^{-11}$. More particular, $c_{(32)R}$ gives the dominant contribution to  Br$(\tau\rightarrow \mu\gamma)<4.4\times 10^{-8}$, with $|c^{H_2}_{(32)R}|\gg |c^{H_1}_{(32)R}|$.  In contrast,  the remaining cLFV decays have some common properties that  $|c_{(ab)R}|>|c_{(ba)R}|$ with   $a<b$,   $|c^{H_k}_{(ab)R}|\gg |c_{(ab)R}|$, and the huge destructive correlation between two charged Higgs boson contributions. They are very important to guarantee small Br$(\tau \rightarrow e\gamma)$ and Br$(\mu  \rightarrow e\gamma)$.  On the other hand, they allow large and/or constructive $c^{H_k}_{(22)R}$, which are the dominant contributions resulting in large $\De a^{331\mathrm{ISS}}_{\mu}\ge 10^{-9}$.

The above properties are also true for the allowed region of the parameter space given in Table~\ref{t_range1free}.  They are summarized in Table~\ref{t_Rxy} through the quantities defined in Eqs.~\eqref{eq_RabX} and \eqref{eq_Rmab}.
\begin{table}[ht]
	\centering
	\begin{tabular}{ccccccccccccccccc}
		\hline
		& $	R^W_{22}$& $R^{H_1}_{22}$& $R^{H_2}_{22}$ & $	R^W_{12}$& $	R^Y_{12}$& $R^{H_1}_{12}$& $R^{H_2}_{12}$& $R^{-}_{12}$&$	R^W_{21}$& $	R^Y_{21}$& $R^{H_1}_{21}$& $R^{H_2}_{21}$& $R^{-}_{21}$& $R^W_{13}$& $	R^Y_{13}$ \\
     Min &0.06  & 0.96 & 0 & 54 & 7 & $\sim10^3$ & $\sim10^3$ & 60 & 0.3 & 0.04 & 34 & 34 & 0.01 & 0.01 & 0 \\
     Max &0.08& 1.11 & 0.1 & $\sim 10^5$ &$\sim 10^4$ &$ \sim10^7$ & $\sim 10^7$ & $\sim 10^6$ & 299 & 37 & $\sim 10^4$ &  $\sim 10^4$ & 336 & 4.& 3 \\
		\hline
		\hline
	 & $R^{H_1}_{13}$ &$R^{H_2}_{13}$& $R^{-}_{13}$	& $R^{H_1}_{31}$& $R^{H_2}_{31}$& $R^{-}_{31}$& $R^W_{23}$ & $R^Y_{23}$ & $R^{H_1}_{23}$ & $R^{H_2}_{23}$& $R^{-}_{23}$ &  $R^{H_1}_{32}$ & $R^{H_2}_{32}$& $R^{-}_{32}$ \\
		Min & 6 &7 & 0.05 & 0.1 & 0.8 &$\simeq1$ & 0.01 & 0.03 & 0.5 & 0 & 0.02 & 0 & 0.96 & 0.998 \\
		Max & $\sim10^3$  & $\sim10^3$& 2.6 &  0.2 & 0.89& $\simeq 1$ & 16.4 & 15.4 & 222 & 255 & 33  & 0.04 & 0.996 & 1.002 \\
		\hline
	\end{tabular}
	\caption{ Correlations between different contributions to $c_{(ab)R}$ with ranges of free parameters given in Table~\ref{t_range1free}, where we denote  $0\simeq R^Y_{(22)R},R^W_{(31)R}, R^Y_{(31)R}, R^W_{(32)R}, R^Y_{(32)R}\leq \mathcal{O}(10^{-3})$ .\label{t_Rxy}}
\end{table}
We can see that $R^{-}_{32}= |\mathrm{Re}[c^{H_2}_{(32)R} +c^{H_1}_{(32)R}]/\mathrm{Re}[c_{(32)R}]| \rightarrow 1$ implies that sum of the two contributions of the charged Higgs bosons to $c_{(32)R}$ is dominant. In addition $R^{H_2}_{32}\equiv |\mathrm{Re}[c^{H_2}_{(32)R}]/ \mathrm{Re}[c_{(32)R}]| \simeq  1$ indicates  that  the  contributions of the charged Higgs boson $H_2$ is dominant, hence the destructive correlation is small. This is not enough to keep the cLFV constraint Br$(\tau \rightarrow \mu\gamma)<4.4\times 10^{-8}$ for larger $\De a_{\mu}>108.5\times 10^{-11}$. All contributions  of the two decays $\mu\rightarrow e\gamma$ and $\tau\rightarrow e\gamma$ do not have properties mentioned here.   In the next discussion, we will show that new destructive contributions from additional singly charged Higgs bosons will relax  the sum of  the contributions from two charged Higgs bosons $H_{1,2}$ to a larger values, while allow both  $\De a^{331\mathrm{ISS}}_{\mu}$ and Br$(\tau \to \mu\gamma)$  satisfying the experimental constraints given in Eqs.~\eqref{eq_damu} and \eqref{eq_ebagaex}.

From the above discussion, we can see that  max$[\Delta a_{\mu}]\simeq 108.5\times 10^{-11}$ predicted by the 331ISS model  comes from the experimental constraints Br$(\tau\rightarrow \mu \gamma)<4.4\times 10^{-8}$, which gets main contributions from $c_{(32)R}$. On the other hand, Br$(\tau \rightarrow e\gamma)$ can reach to zero for large $ \Delta a_{\mu}\ge 10^{-9}$, which is different from the normal behave of these two branching ratios  Br$(\tau\rightarrow \mu \gamma)\sim$Br$(\tau\rightarrow e \gamma)\sim |c_{(23)R}|^2,|c_{(13)R}|^2\gg |c_{(32)R}|^2,|c_{(31)R}|^2$. After some numerical checks, we see that  the this difference is    originated mainly  from  the following property:   each quantity   Br$(\tau\rightarrow \mu \gamma)$ or Br$(\tau\rightarrow e \gamma)$ contains only one type of terms  with a factor  $\frac{m_{\mu}}{m_{\tau}}$  or $\frac{m_{e}}{m_{\tau}}$  appearing in $c_{(32)R}$ or $c_{(31)R}$, respectively. These  terms  are normally suppressed because of many other large terms contained in $c^{H_k}_{(23)R}, c^{H_k}_{(13)R} \gg c^{H_k}_{(32)R},c^{H_k}_{(31)R}$. But when huge destructive correlations between two charged Higgs contributions  and gauge contributions happen, there appears a situation that  $|c_{(13)R}|, |c_{(23)R}| \rightarrow 0$, and also for other normal large terms in $|c_{(31)R}|, |c_{(32)R}| $. Now, the terms with factors $\frac{m_{\mu}}{m_{\tau}}$ and $\frac{m_{e}}{m_{\tau}}$ become significant, leading  to the consequence that Br$(\tau \rightarrow e \gamma)\sim \frac{ m^2_e}{m^2_{\tau}} $ can be close to 0, while Br$(\tau \rightarrow \mu \gamma)\sim \frac{ m^2_\mu}{m^2_{\tau}}\gg $Br$(\tau \rightarrow e \gamma)$. It is reasonable to think that the  terms with factor $\frac{m_{\mu}}{m_{\tau}}$ and  $\Delta a^{331\mathrm{ISS}}_{\mu}$ get similar contributions relating to  $\mu$, hence   both of them   must be large if  $\Delta a^{331\mathrm{ISS}}_{\mu}$ is required  to be large in order to reach the experimental constraints.  Our explanation is confirmed by a numerical check, where we change $m_{\mu} \to m_{e}$ in only the formula of $c_{(32)R}$. We saw that Br$(\tau \to \mu \gamma, e\gamma)$ can reach  small values Br$(\tau \to \mu \gamma,e\gamma)<10^{-9}$ with  $\Delta a^{331\mathrm{ISS}}_{\mu}>125\times 10^{-11}$.  Other numerical checks also show that the lower bound of Br$(\tau \rightarrow \mu\gamma)$  depends strictly  on the lepton mixing matrix $U_{\mathrm{PMNS}}$, which is the only cLFV source in the 331ISS model.   First,  the case of  large $\tau-e$ mixing inputs  $s^2_{13}=s^2_{23}=0.547$ can give  large  $\Delta a^{331\mathrm{ISS}}_{\mu}>115\times 10^{-11}$  and both small Br$(\tau \rightarrow \mu \gamma, e\gamma)\to 0$.   Second, the small input $s^2_{23}=0.0216$ and the large input  $s^2_{13}=0.547$ will   result in that max$[\Delta a^{331}_{\mu}]\simeq 90\times 10^{-11}$. In both cases,  max$[\Delta a^{331\mathrm{ISS}}_{\mu}]$ is  still constrained by Br$(\tau \to \mu \gamma)<4.4\times 10^{-8}$. In conclusion, the regions of the parameter space giving  max$[a^{331\mathrm{ISS}}_{\mu}]$ allows  all  small $c_{(ab)R}$ except the terms with factor  $\frac{m_{\mu}}{m_{\tau}}$ in $c_{(32)R}$.

 \subsection{ New contributions from additional singly charged Higgs bosons $h^\pm_3$}
\begin{figure}[ht]
 	\centering
 	\begin{tabular}{cc}
 		\includegraphics[width=7cm]{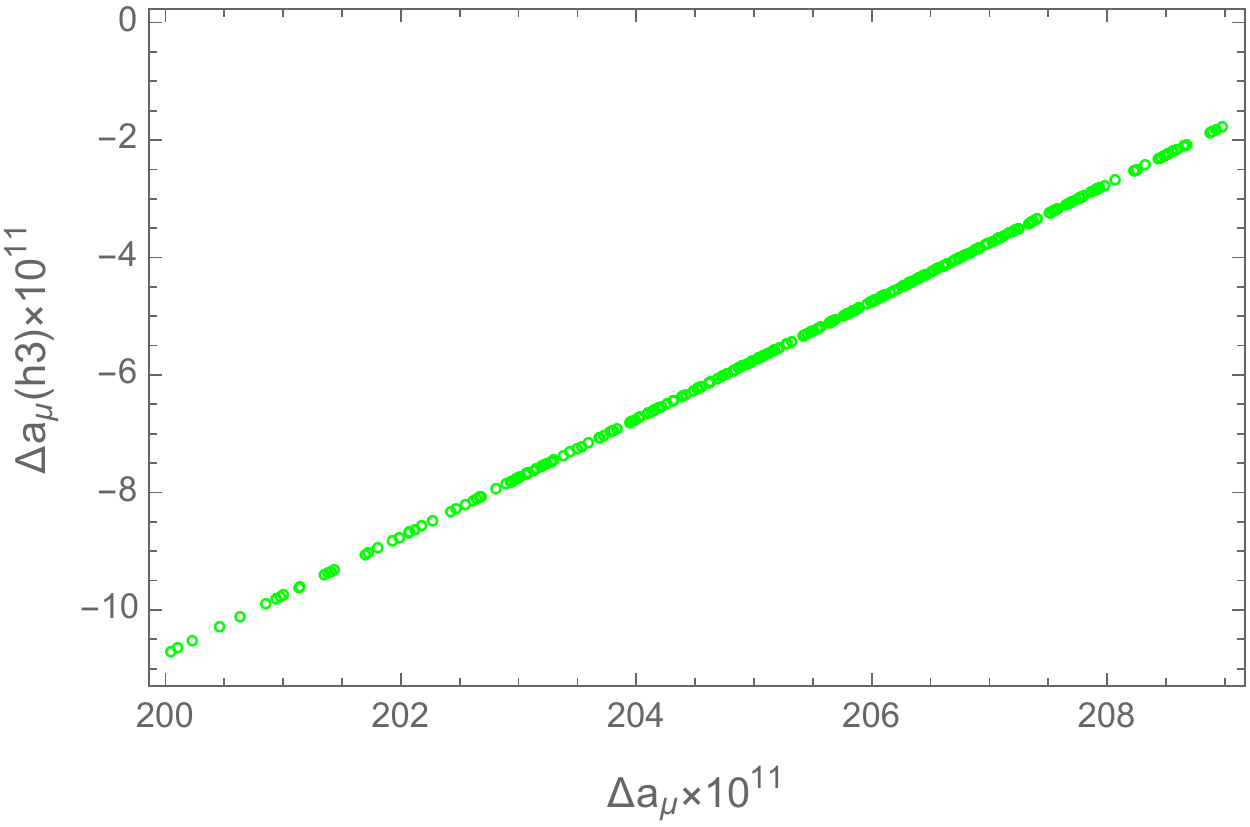}	&	
 		\includegraphics[width=7cm]{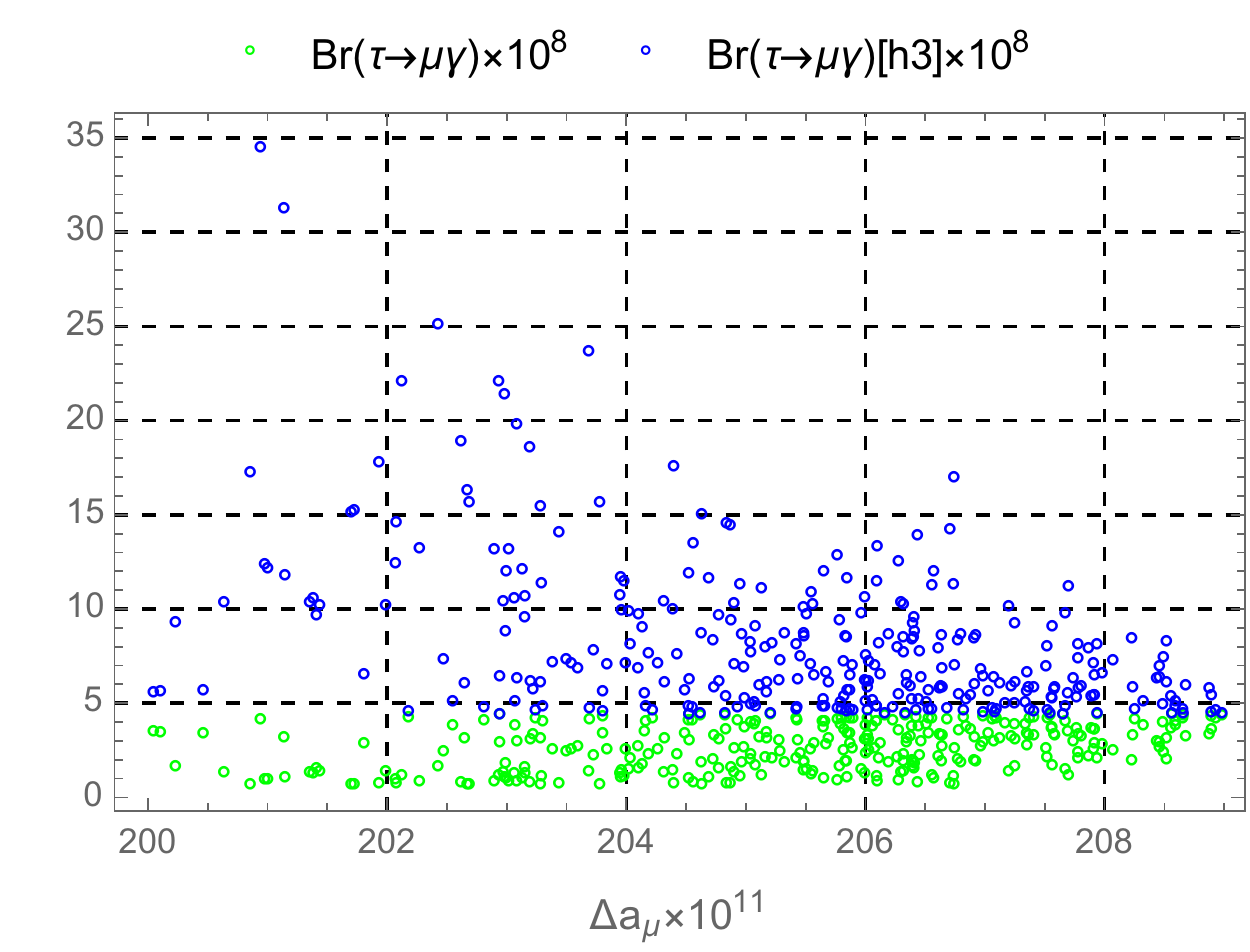}\\
 	\end{tabular}
 	\caption{  Correlations between  $\Delta a_{\mu}\equiv \Delta a^{331\mathrm{ISS}}_{\mu}$ with   $\De a_{\mu}(h_3)$ and Br$(\tau \rightarrow \mu\gamma)[h_3]$. }\label{fig_amu1sigma}
 \end{figure}

 Adding contributions of the  new singly charged  Higgs boson, \ the allowed values of $\De a_{\mu}\equiv \De a^{331\mathrm{ISS}}_{\mu}\ge 192\times 10^{-11} $ corresponding to the lower bound  of the  $1\sigma$ confidence level  are explained  successfully, see  an illustration shown in Fig.~\ref{fig_amu1sigma},
 where $\De a_{\mu}(h_3)$ and Br$(\tau \rightarrow \mu\gamma)[h_3]$ show the  respective one-loop contributions from only $h^\pm_3$ to $\De a_{\mu}$ and Br$(\tau \rightarrow \mu\gamma)$, which are defined as follows:
 \begin{align}
 	\Delta a_{\mu} [h_3]&= -\frac{4 m^2_{\mu}}{e} \mathrm{Re}[c^{h_3}_{(22)R}] ,\crn
 	\mathrm{Br}(e_b\rightarrow e_a \gamma)[h_3]&=\frac{48\pi^2}{ G_F^2} \left( \left|c^{h_3}_{(ab)R}\right|^2 +\left|c^{h_3}_{(ba)R}\right|^2\right) \mathrm{Br}(e_b\rightarrow e_a\overline{\nu_a}\nu_b).
 \end{align}
 The corresponding benchmark is calculated numerically with  30 digits of precision number. The numerical values of the  free parameters are
\begin{align}
	\label{eq_allwedPoint}
	k_{11} &\simeq  -19.19,\;
	k_{22} \simeq -94.53,\;
	k_{33} \simeq  428.75,\;
	k_{12} \simeq -89.46,\crn
	k_{13} &\simeq  29.47,\;
	k_{23} \simeq  -211.84,\;
	k_{21} \simeq 60.09,\;
	k_{31} \simeq -262.44,\;
	k_{32} \simeq 30.53,\crn
	t_{\beta}&= 49.86, \; z= 1169\;\mathrm{GeV},\; m_{H_1}= 657.1  \;\mathrm{GeV}, \;  m_{H_2}=   734  \;\mathrm{GeV}.
	%
\end{align}
In this case, the heavy neutrino masses are $m_{n_4}=m_{n_5}= 137.2$ GeV, $m_{n_6}=m_{n_7}=4709.4$ GeV, $m_{n_8}=m_{n_9}=11958$ GeV.
For simplicity we assume that $Y^3_{11}=Y^3_{12}=Y^3_{21}=Y^3_{13}=Y^3_{31}=0$, therefore the  contribution from  $h_3$ does not change the two  cLFV decays  Br$(\mu\rightarrow e\gamma)\simeq 3.93\times 10^{-13}$ and Br$(\tau\rightarrow e\gamma)\simeq 1.11\times 10^{-8}$.  They always satisfy the experimental data. The non-zero Yukawa couplings are scanned in the ranges $Y^3_{ab}\in [-3.5,3.5]$ that satisfy the perturbative limit.  This results in the following allowed range of the charged Higgs boson  mass  $ 500\; \mathrm{GeV} \le m_{h_3}\le \;1158 \mathrm{GeV}$. Numerical values of $c_{(ab)R}$ is shown in Table~\ref{t_largeDeamuh3}.
\begin{table}[ht]
	\begin{tabular}{cccccccc}
		\hline
		Notations	&	$c^W_{(ab)R} -c^{W,\mathrm{SM}}_{(ab)R}$ &$c^Y_{(ab)R}$ & $c^{H_1}_{(ab)R}$ & $c^{H_2}_{(ab)R}$ & $c^{h_3}_{(ab)R}$ & $c_{(ab)R}$ & Process \\
		\hline
		\hline
		$\Delta a_{\mu}:\;c_{(22)R}\times 10^{10}$& 5.3 &-0.386 & -211. & 61.1 &  3.7  & -141.1  &  $  \Delta a_{\mu} =20.5 \times 10^{-10}$  \\
		$\mu\rightarrow e\gamma:\;c_{(12)R}\times 10^{13}$& 449.16 & 61.536 & -75957. & 75443. & 0 & -2.5234   &  $\; \mathrm{Br}(12R) = 2.2174 \times 10^{-13}$   \\
		$\mu\rightarrow e\gamma:\; c_{(21)R}\times 10^{13}$& 2.1388 & 0.29303 & -361.70 & 357.43  & 0 & -1.8329  & $\; \mathrm{Br}(21R) = 1.1699 \times 10^{-13}$   \\
		$\tau\rightarrow e\gamma:\; c_{(13)R}\times 10^{10}$&  -0.00510 & 0.0540 & 4.25 & -2.96 &0 & 1.34 &  $\; \mathrm{Br}(13R) = 111.\times 10^{-10}$    \\
		$\tau\rightarrow e\gamma:\; c_{(31)R}\times 10^{10}$&$\sim 0$ & $\sim 0$& 0.00120 & 0.0664  & 0 & 0.0676 &  $\; \mathrm{Br}(31R) = 0.284\times 10^{-10}$    \\
		$\tau\rightarrow \mu \gamma:\; c_{(23)R}\times 10^{10}$&  -0.00721 & -0.0445 & 1.20 & -2.51 &  0.164 & -1.20 & $\; \mathrm{Br}(23R) =86.7 \times 10^{-10}$     \\
		$\tau\rightarrow \mu \gamma:\; c_{(32)R}\times 10^{10}$& -0.000426 & -0.00263 & 0.0708 & -5.18 &  2.77& -2.33 & $\; \mathrm{Br}(32R) = 330. \times 10^{-10}$    \\
		\hline
	\end{tabular}
	\caption{ Particular contributions $c^X_{(ab)R}[\mathrm{GeV}^{-2}]$ to $\Delta a_{\mu}$ and Br$(e_b\rightarrow e_a\gamma)$ with the free parameters shown in Eq.~\eqref{eq_allwedPoint}. The last column shows values of $\Delta a_{\mu}$ and  Br$(e_b\rightarrow e_a\gamma)$.  \label{t_largeDeamuh3}}
\end{table}

The numerical results shown in Fig.~\ref{fig_amu1sigma} have some interesting properties.  In the left panel,  the contributions from $h^\pm_3$ to $\De a_{\mu}$ are always negative, but much smaller than the total one: $0<-\De a_{\mu}(h^\pm_3)\le 1.5\times 10^{-10}\ll 200\times 10^{-11}\sim \De a_{\mu}$.   On the other hand, the one-loop contributions $c^{h_3}_{(32)R}$ and  $c^{H_2}_{(32)R}$ have  the same order, but     opposite signs. Therefore, the total  $|c_{(32)R}|$ is small enough to guarantee that Br$(\tau \rightarrow \mu\gamma)<4.4\times 10^{-8}$. This is reason why in the right panel, we see that  $|c_{(32)R}|<|c^{h_3}_{(32)R}|$, i.e. Br$(\tau \rightarrow \mu\gamma)<$  Br$(\tau \rightarrow \mu\gamma)[h_3]$ may happen. More specifically, this property can be seen  from a particular numerical illustration presented in Table~\ref{t_largeDeamuh3}.   We can see a property that  $|c_{(22)R}|\gg |c^{h_3}_{(22)R}|\sim |c^{h_3}_{(32)R}|\sim  |c^{H_2}_{(32)R}|\sim |c_{(32)R}|$, which  explains why the contributions from $h_3$ affect strongly   Br$(\tau\rightarrow \mu\gamma)$ but weakly $\Delta a_{\mu}$.

\begin{figure}[ht]
	\centering
	\begin{tabular}{cc}
		\includegraphics[width=7cm]{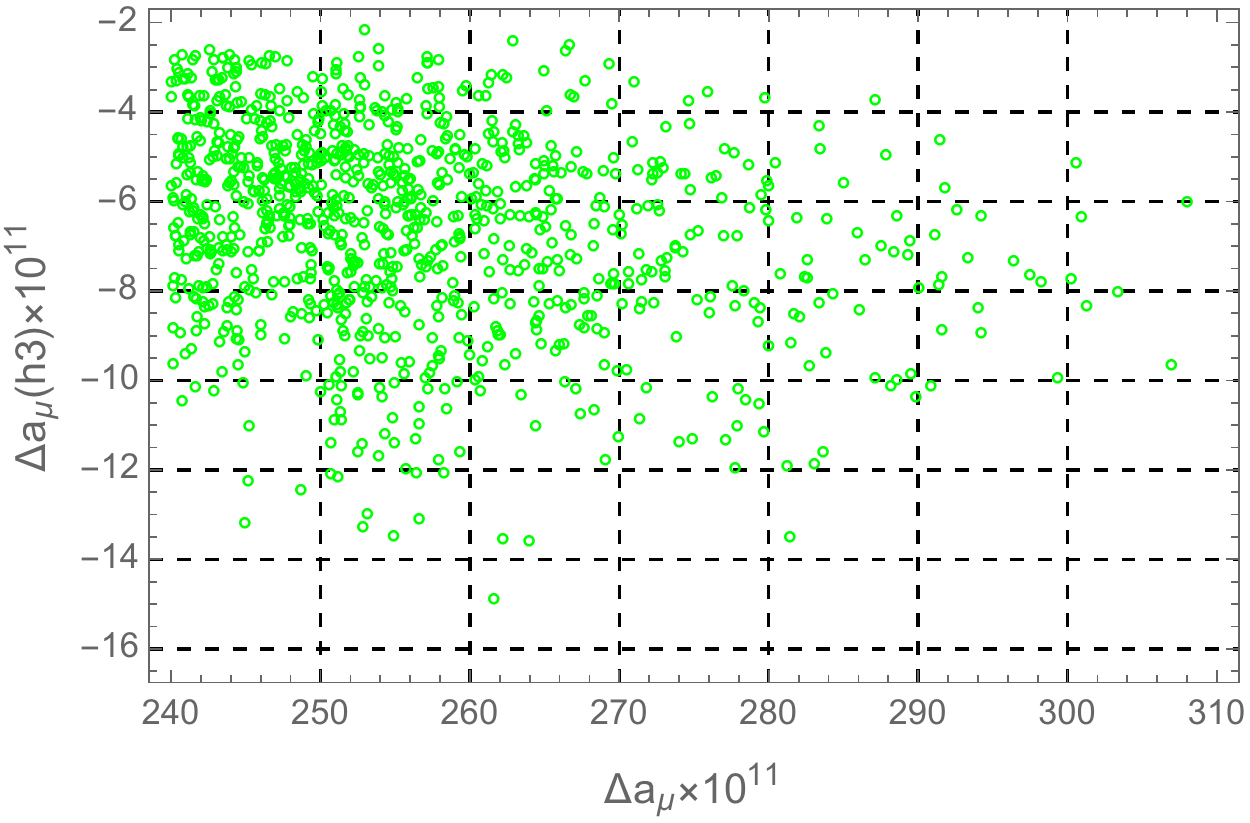}	&	
		\includegraphics[width=7cm]{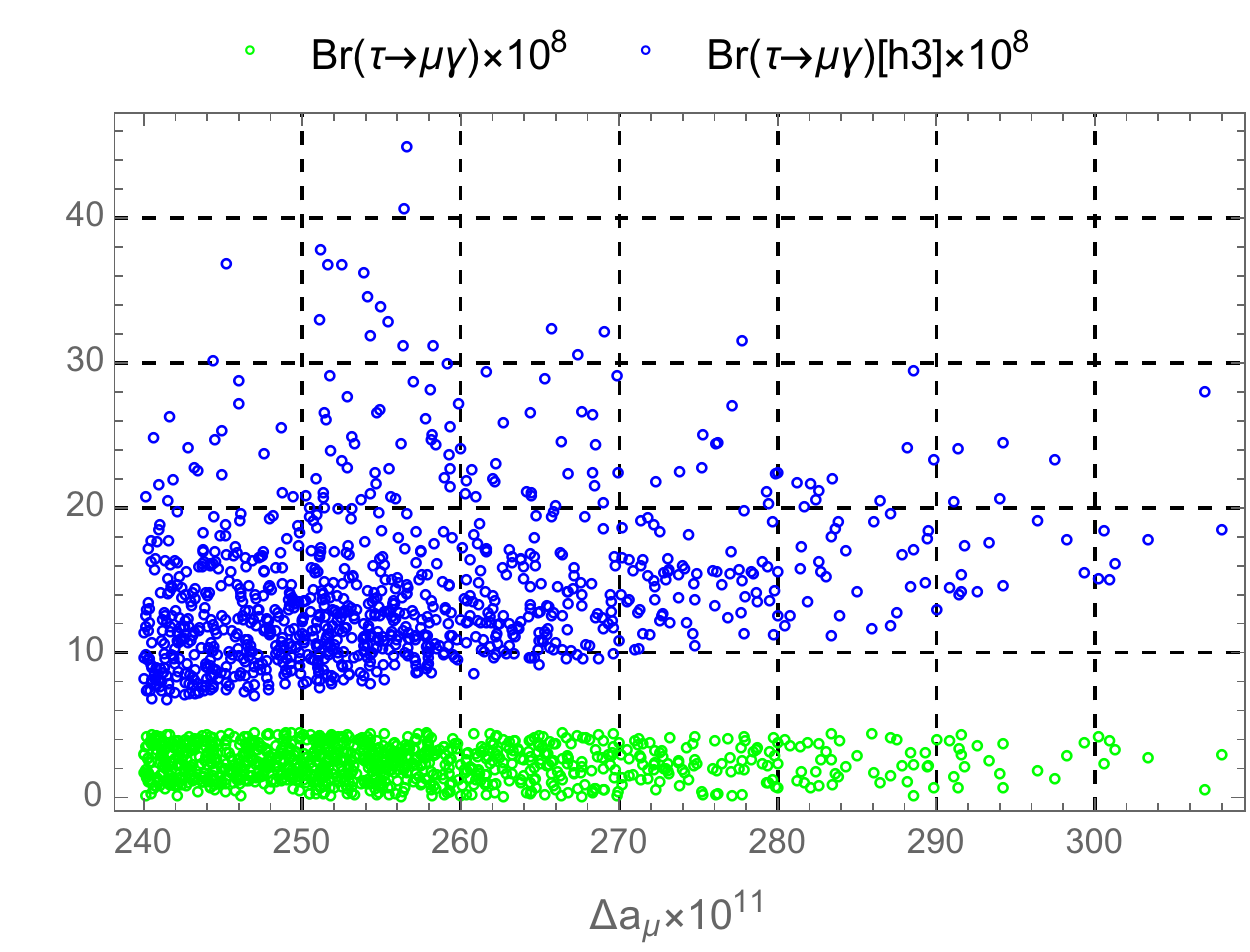}\\
	\end{tabular}
	\caption{  Correlations between  $\Delta a_{\mu}\equiv \Delta a^{331\mathrm{ISS}}_{\mu}\ge 240\times 10^{-11}$ with   $\De a_{\mu}(h_3)$ and Br$(\tau \rightarrow \mu\gamma)[h_3]$.  }\label{fig_amularge}
\end{figure}

The allowed regions of parameters allowing $\Delta a_{\mu}^{331\mathrm{ISS}}$ around the value $200\times 10^{-11}$ can be found easily in the ranges given in Eq.~\eqref{eq_scanNO1}. The allowed regions with larger $\Delta a_{\mu}^{331\mathrm{ISS}}$ are shown in Fig.~\ref{fig_amularge}, where  charged Higgs masses have to be smaller than 600 GeV.
It is noted that  large  $\Delta a_{\mu}^{331\mathrm{ISS}}>300\times 10^{-11}$ require  light charged Higgs boson masses $ m_{H_{1}}\to 500$  GeV, $z\rightarrow 1223$ GeV, and large $t_{\beta}\to 60$. The  region of parameter space corresponding to the Fig~\ref{fig_amularge} is:
\begin{align}
	\label{eq_allowedrange300}
	k_{11} &\in \left[-21.77,\; -17.84 \right],\;
	k_{22} \in \left[ -101.9,\; -93.76 \right],\;
	k_{33} \in \left[420.1,\; 429.4 \right],\crn
	k_{12} &\in \left[ -96.22,\; -88.92 \right],\;
	k_{13} \in \left[26.95,\;31.12 \right],\;
	k_{23} \in \left[ -220.2,\; -210.4 \right],\crn
	k_{21} &\in \left[ 59.19,\; 66.55 \right],\;
	k_{31} \in \left[-268.6,\;- 262.9 \right],\;
	k_{32}\in \left[ 25.35,\; 33.64 \right],\crn
	t_{\beta}&\in \left[ 41.68,\; 59.97 \right], \; z \in \left[1051,\; 1223 \right] \;\mathrm{GeV},\; m_{H_1}\in \left[ 500.6,\; 631.3 \right]  \;\mathrm{GeV},  \crn
	m_{H_2}&\in \left[571.3,\; 703.8 \right]  \;\mathrm{GeV},\; m_{h_3} \in \left[ 500.5,\; 778.6 \right]\;\mathrm{GeV},  \;  |Y_{22}|\in \left[0.11,\;3.49 \right], \crn
	|Y_{23}| &\in \left[ 0.51,\; 3.5 \right], \;   |Y_{32}|\in \left[ 0.06,\; 3.49 \right], \;  |Y_{33}|\in \left[0.009,\; 3.5 \right].
\end{align}
The heavy neutrino masses are in the following ranges: $m_{n_4}=m_{n_5} \in \left[ 109.2,\; 172.3 \right]$ GeV, $m_{n_6}=m_{n_7}\in \left[ 3.66,\; 5.87\right]$ TeV, $m_{n_8}=m_{n_9} \in \left[ 8.99,\; 14.92\right]$ TeV.  The  cLFV branching ratios are in the following ranges: Br$(\mu\rightarrow e\gamma)\times 10^{13} \in [5.8\times 10^{-16},\; 4.2\times 10^{-13}]$, Br$(\tau\rightarrow e\gamma) \in [4\times 10^{-11},\; 3.3\times 10^{-8}]$, and  Br$(\tau\rightarrow \mu\gamma) \in [1.6\times 10^{-12},\; 4.4\times 10^{-8}]$.

\section{\label{sec:conclusion}Conclusion}
In this work, we have pointed out that the one of the versions of the 3-3-1RN model, namely the 331ISS model,   can predict large  values of  $\De a_{\mu}\simeq 108\times 10^{-11}$   under the recent constraint of all cLFV decays $e_b\rightarrow e_a\gamma$.  This large value corresponds to  the upper bound Br$(\tau\rightarrow \mu\gamma)\simeq 4.4\times 10^{-8}$, while the two remaining decay branching ratios  are still well below the recent experimental constraints. This model predicts the existence of the two charged Higgs bosons which can give large    contributions  of the order $\mathcal{O}(10^{-9})-\mathcal{O}(10^{-8})$ to the $\De a_{\mu}$, so that it  can reach the maximal values around $10^{-9}$,  which  is still much smaller than the allowed values given by the recent experimental data. On the other hand,  the two other charged Higgs bosons contributions to Br($e_b\rightarrow e_a\gamma$) will be at the orders of     $\mathcal{O}(10^{-10})-\mathcal{O}(10^{-9}) [\mathrm{GeV}^{-2}]$. But the huge destructive correlations can happen between these contributions, leading to a small values of  Br$(e_b \rightarrow e_a\gamma)$. Although the model contains many free parameters, maybe the  antisymmetry of the  Dirac mass matrix $m_D$ does not allow large destruction enough to keep the Br$(\tau \rightarrow \mu\gamma)$ below the experimental constraint, while allow large $\Delta a_{\mu}^{331\mathrm{ISS}}\ge 192\times 10^{-11}$. The model needs to include an additional singly charged Higgs boson so that  all experimental data of  $\De a_{\mu}$ and the cLFV decays  can be explained simultaneously.  As a consequence, all of the  cLFV  decays $e_b\rightarrow e_a\gamma$  are predicted that their branching ratios can be large closely  the recent experimental bounds.  Therefore, our model can also explain simultaneously all cLFV decays  $e_b\rightarrow e_a\gamma$ once they are observed by upcoming experiments.
\section*{Acknowledgments}
We are grateful to Prof. Martin Hoferichter for introducing us the latest result of the  SM prediction of  $\De a_{\mu}$. We  thank Prof. Hidezumi Terazawa, Dr. Wen Yin, and  Dr.  Pengxuan Zhu for their  useful information. We would like to express our sincere gratitude to the referee for correcting the electron mass in the original draft, leading to the new numerical illustration in the new version. This research is funded by Vietnam  National Foundation for Science and Technology Development (NAFOSTED) under grant number  103.01-2018.331.

\appendix

\end{document}